\documentclass[final,5p,times,twocolumn,authoryear]{elsarticle}
\usepackage{graphicx}	
\usepackage{amsmath}	
\usepackage{amssymb}
\usepackage{lipsum}
\usepackage[utf8]{inputenc} 
 
\usepackage{bm}		

\usepackage{color}
\usepackage{pdflscape}
\usepackage{lscape}
\usepackage{mathtext}
\usepackage{float}
\usepackage{textcomp}
\usepackage{rotating}
\usepackage{dcolumn}
\usepackage{tabularx}
\usepackage{pdflscape} 
\usepackage{hyperref}
\usepackage{textcomp}

\def\flux{\rm erg~s$^{-1}$~cm$^{-2}$}
\def\lum{\rm erg~s$^{-1}$}

\def\lumU{\rm erg~s^{-1}}
\def\lumdU{\rm erg~s^{-1}~Hz^{-1}}
\def\arcsec{$^{\prime\prime}$}
\def\arcmin{$^{\prime}$}

\newcommand{\Lx}{L_{\rm 2keV}}

\newcommand{\Fx}{F_{\rm X}}

\newcommand{\s}[1]{\sigma_{\rm{#1}}}
\newcommand{\Var}{{\rm Var}}
\newcommand{\Varj}[1]{\hat{{\rm Var}}_{#1}}
\newcommand{\Sfx}{\widetilde{SF}^2_{\rm X}}
\newcommand{\Sfxi}{\widetilde{SF}^2_{\rm X,i}}
\newcommand{\Sf}{\widetilde{SF}^2}
\newcommand{\SfO}{\hat{SF}^2}
\newcommand{\Sfj}[1]{\hat{SF}^2_{#1}}
\newcommand{\Sigx}{\sigma^2_{\rm X}}
\newcommand{\Siguv}{\sigma^2_{\rm UV}}
\newcommand{\Sigtotx}{\sigma^2_{{\rm totX}}}
\newcommand{\Sigtotuv}{\sigma^2_{{\rm totUV}}}
\newcommand{\Sigmx}{\sigma^2_{{\rm mX}}}
\newcommand{\Sigmuv}{\sigma^2_{{\rm mUV}}}
\newcommand{\Sigvarx}{\sigma^2_{{\rm varX}}}
\newcommand{\Sigvaruv}{\sigma^2_{{\rm varUV}}}
\newcommand{\Sigintx}{\sigma^2_{{\rm intX}}}
\newcommand{\Sigintuv}{\sigma^2_{{\rm intUV}}}

\newcommand{\Luv}{L_{2500}}
\newcommand{\lx}{l_{\rm X}}
\newcommand{\luv}{l_{\rm UV}}

\newcommand{\plx}{\tilde l_{\rm X}}
\newcommand{\pluv}{\tilde l_{\rm UV}}
\newcommand{\pLx}{\tilde L_{\rm X}}
\newcommand{\pLuv}{\tilde L_{\rm UV}}

\newcommand{\pLF}{\widetilde{LF}}
\newcommand{\npLF}{\widetilde{NLF}}
\newcommand{\pLFuv}{\widetilde{LF}_{UV}}

\newcommand{\Tc}{\Delta T_{\rm c}}

\newcommand{\srg}{{\it SRG}}
\newcommand{\xmm}{{\it XMM-Newton}}

\def\wid{6mm}

\journal{High Energy Astrophysics}

\begin{document}

\begin{frontmatter}


\title{SRG/eROSITA–SDSS view on the relation between X-ray and UV emission for quasars}
\author[first,second]{Sergey Prokhorenko}
\ead{sprokhorenko@cosmos.ru}
\author[first,second]{Sergey Sazonov}
\author[first,third]{Marat Gilfanov}
\author[fourth]{Sergey Balashev}
\author[first]{Alexander Meshcheryakov}
\author[fourth]{Alexander Ivanchik}
\author[sixth,seventh]{Ilfan Bikmaev}
\author[first,third,eigth]{Rashid Sunyaev}

\affiliation[first]{organization={Space Research Institute of the Russian Academy of Sciences},
            addressline={Profsoyuznaya Str. 84/32}, 
            city={Moscow},
            postcode={117997}, 
            country={Russia}}
\affiliation[second]{organization={National Research University Higher School of Economics},
            addressline={Pokrovsky Bulvar 11}, 
            city={Moscow},
            postcode={101990}, 
            country={Russia}}
\affiliation[third]{organization={Max-Planck-Institut f\"{u}r Astrophysik},
            addressline={Karl-Schwarzschild-Str. 1}, 
            city={Garching},
            postcode={D-85741}, 
            country={Germany}}
\affiliation[fourth]{organization={Ioffe Institute},
            addressline={Politeknicheskaya str. 26}, 
            city={St Petersburg},
            postcode={194021}, 
            country={Russia}}
\affiliation[sixth]{organization={Kazan Federal University},
            addressline={Kremlevskaya str. 18}, 
            city={Kazan},
            postcode={420008}, 
            country={Russia}}
\affiliation[seventh]{organization={Academy of Sciences of Tatarstan},
            addressline={Baumana Str. 20}, 
            city={Kazan},
            postcode={420111}, 
            country={Russia}}      
\affiliation[eigth]{organization={Institute for Advanced Study},
            addressline={1st Einstein Drive}, 
            city={Princeton},
            postcode={08540}, 
            state={New Jersey},
            country={USA}}

\begin{abstract}
Motivated by the idea of using quasars as standardizable candles for cosmology, we examine the relation between X-ray (at 2\,keV, $\Lx$) and ultraviolet (at 2500\,\AA, $\Luv$) monochromatic luminosities of quasars using a sample of 2414 X-ray sources from the \srg/eROSITA all-sky survey cross-matched with the Sloan Digital Sky Survey data release 16 quasar catalogue (SDSS DR16Q), at spectroscopic redshifts between 0.5 and 2.5. These objects are bright both in X-rays and in the optical, so that the sample is characterized by nearly 100\% statistical completeness. We have developed a new method for determining the $\Lx$--$\Luv$ relation, which consistently takes into account (i) X-ray and UV flux limited object selection, (ii) X-ray and UV variability of quasars, and (iii) the decreasing space density of quasars with increasing luminosity. Assuming a linear relation between $\lx\equiv\log{\frac{\Lx}{\lumdU}}$ and $\luv\equiv\log{\frac{\Luv}{\lumdU}}$, we find the slope, $\gamma=0.69\pm0.02$ (hereafter all uncertainties are quoted at the 68\% confidence level), and normalization, $\lx=26.47\pm0.02$ at $\luv=30.5$, of the $\Lx$ ($\Luv$) dependence. These values are not substantially different from the results of previous studies. A key novel aspect of our work is allowance for intrinsic scatter (which adds to the dispersion induced by quasar variability and flux measurement uncertainties) of the $\Lx$--$\Luv$ relation in both variables, i.e. in X-ray and UV luminosity. The intrinsic X-ray scatter ($\Sigintx=0.063\pm0.005$) strongly dominates over the UV one ($\Sigintuv=0.002^{+0.003}_{-0.002}$). Further studies should seek to explain this behaviour in terms of accretion onto supermassive black holes and orientation of quasars with respect to the observer. 

\end{abstract}


\end{frontmatter}

\section{Introduction}
\label{s:introduction}

Active galactic nuclei (AGN) are extremely diverse objects as regards their luminosity, spectral properties, radio-loudness, variability, polarization etc. (see \citealt{padovani2017} for a review). However, this diversity is predominantly driven by just a few physical parameters related to accretion of matter onto the central supermassive black hole (SMBH), namely its mass, spin and accretion rate, with a presumably lesser role played by the properties of the host galaxy. This should cause different AGN observable properties to correlate with each other, and a number of such correlations have indeed been found. 

One of the most notable and reliably established relations is between the X-ray and ultraviolet (UV) luminosities, usually expressed in terms of monochromatic luminosities (per frequency) at rest-frame 2\,keV and 2500\,\AA, which we will hereafter refer to as $\Lx$ and $\Luv$, respectively. Its existence was first suggested in the 1970s--1980s \citep{tananbaum1979,Zamorani1981,Avni1982,avni1986} and has then been validated in many studies using progressively larger AGN samples. Most authors conclude that the relation between $\Lx$ and $\Luv$ is nonlinear, namely that $\Lx/\Luv$ decreases with increasing $\Luv$ (e.g. \citealt{lusso2010,bisogni2021chandra,dainotti2022,signorini2024quasars,rankine2024,Sacchi2025LxLuv,chira2025}). This implies a decreasing contribution of X-ray emission (presumably produced by Comptonization in a hot corona of the accretion disc) to the AGN bolometric luminosity as the latter increases. The 2500\,\AA\ luminosity, which can be conveniently measured by ground-based spectrographs for objects located at redshifts between approximately 0.5 and 3, is believed to be a good tracer of thermal emission from the inner regions of the accretion disc (e.g. \citealt{Jin2023}). 

The $\Lx$--$\Luv$ relation is not only important for exploring the internal structure of AGN and the physics of accretion onto black holes but also for cosmology, as it potentially provides an opportunity to use quasars (i.e. luminous and distant AGN), as standardizable candles for building the cosmic distance ladder, even at high redshifts \citep{risaliti2015hubble,Lusso_2019T,lusso2020quasars,Khadka_2020,Li_2022z,Lenart_2023}. However, this can only be fulfilled if this relation is indeed tight and nonlinear. Otherwise it would be impossible to deduce the absolute distance to a quasar by measuring $\Lx$ and $\Luv$. In reality, previous studies have reported significantly diverse values of the slope of the $\Lx (\Luv$) dependence\footnote{Virtually all authors have adopted, for simplicity, a power-law form for this relation.} \citep{Vignali_2003,strateva2005,steffen2006x,bisogni2021chandra,rankine2024} and found a substantial scatter in the correlation: $\Lx/\Luv$ varies by a factor of $\sim 2$ for a given luminosity \citep{lusso2016tight,Lusso2017}. As a result, the practical usability of quasars for cosmology is still under question, urging further investigation. The precision of previous studies of the X-ray--UV luminosity relation was often limited by the size of the AGN samples used. In addition, it is important to carefully account for selection effects, both in the X-ray and optical--UV bands. Another important issue, as we demonstrate below, is to properly define what is actually meant by the X-ray--UV luminosity relation and its scatter, since different definitions will lead to different results. 

Here, we use a new large and well defined sample of AGN (which we hereafter refer to as quasars) obtained by cross-matching the catalogue of X-ray sources detected during the all-sky survey by the eROSITA telescope \citep{predehl2021} aboard the \srg\ observatory \citep{sunyaev2021} in the Eastern Galactic half of the sky with the catalogue of quasars observed during the SDSS optical spectroscopic survey \citep{lyke2020}. This sample comprises about two and a half thousand quasars and effectively spans two decades in luminosity. This allows us to conduct one of the most precise investigations of the $\Lx$--$\Luv$ relation thus far. To this end, we have developed a novel method that self-consistently takes X-ray and optical/UV selection effects into account and, more importantly, for the first time allows for the presence of intrinsic scatter of the $\Lx$--$\Luv$ relation in both variables, i.e. in X-ray and UV luminosity. Throughout the paper, we use a flat ${\rm\Lambda CDM}$-cosmology model with $H_0=70\,{\rm km\,s^{-1}\,Mpc^{-1}}$, $\Omega_{M}=0.3$, and $\Omega_\Lambda=0.7$.

\section{SRG/eROSITA--SDSS sample}
\label{s:sample}

Our sample is composed of spectroscopically confirmed quasars from the 16th data release of the Sloan Digital Sky Survey (SDSS) Quasar Catalogue (SDSS DR16Q, \citealt{lyke2020}) that were detected by eROSITA during the \srg\ all-sky survey in the $0^\circ<l<180^\circ$ hemisphere. Specifically, we use the eROSITA catalogue of point sources detected on the summed map of the first five all-sky surveys\footnote{The fifth eROSITA survey was interrupted on February 26th, 2022 and has not resumed yet, so that for $\sim 60$\% of the sky the data of only four surveys are available.} in the 0.3--2.3\,keV energy band. The SDSS DR16Q catalogue covers 9376 square degrees in the sky, of which 7994 are at $0^\circ<l<180^\circ$. 

\subsection{Raw sample}
\label{s:rawsample}

To minimize systematic uncertainties and biases associated with absorption in the interstellar medium of the Galaxy, we excluded sky regions with Galactic neutral hydrogen column density \citep{bekhti2016hi4pi} $N_{\rm H}>4\times 10^{20}$\,cm$^{-2}$. The attenuation in the $g$ band associated with Galactic dust \citep{schlegel1998maps,schlafly2011measuring} is less than 0.5\,mag for the remaining sky area of 4965 square degrees. In addition, we selected for our analysis only those regions that had been covered by spectroscopy during all three of the SDSS-I/II, SDSS-III and SDSS-IV survey stages. Given the complex optical/infrared colour and variability based selection of quasar candidates for SDSS spectroscopy (\citealt{berk2005,richards2006,dawson2013baryon,myers2015sdss}), more than 98\% of quasars brighter than $m_g=19$ at redshifts between 0.5 and 2.5 (the range of our interest) are expected to have been covered by spectroscopy in these regions (as we explicitly demonstrate in \ref{appendix:A}), whereas in regions that were covered by only one or two SDSS quasar programs (I/II, III or IV), there can be a deficit of either dim or bright quasars, which we would like to avoid. This further reduces the area of our study to 2858 square degrees, which finally defines our \srg/eROSITA--SDSS footprint. 

We cross-matched the SDSS and eROSITA catalogues using the X-ray position uncertainties, $r98$. This is the radius of the circle within which the source is located with a probability of 98\%, according to eROSITA. We limited our sample of quasars in both the X-ray and optical flux, as described below. Upon application of the flux limits, the median value of $r98$ is 5\arcsec, and 90\% of objects have $r98<7$\arcsec. For 11 eROSITA sources, there are two SDSS quasars within $r98$. We exclude these ambiguous cross-matches from our sample.

In the X-ray domain, our sample is limited by the Galactic absorption corrected flux in the observed 0.3--2.3\,keV energy band (which corresponds to a range between $\sim 0.5$ and $\sim 5$\,keV in the quasar rest-frame for a typical redshift $z\sim 1$ of our objects): $F_{\rm X}>F_{{\rm X,min}}=6\times 10^{-14}$\,erg\,s$^{-1}$\,cm$^{-2}$. At these X-ray fluxes, we have at least 20 eROSITA detector counts from each source (apart from two sources in the final sample, which have 19.9 and 17.0 background subtracted counts) and thus deal with nearly Gaussian statistics of measured fluxes. 

We estimated source X-ray fluxes from the count rates measured by eROSITA, assuming an intrinsic power-law spectrum with a photon index $\Gamma=2$, modified by absorption in the Galactic interstellar medium. To this end, we used the \emph{XSPEC} software (version 12.12.0, \citealt{Arnaud}), the Galactic interstellar absorption model of \cite{Wilms} (\textsc{model TBabs * powerlaw} command in \emph{XSPEC}) with solar abundances of elements, and the hydrogen map from the HI4PI survey \citep{bekhti2016hi4pi}. The adopted slope $\Gamma=2$ may be regarded as a typical effective spectral slope for type I AGN in the $\sim 0.5$ to $\sim 5$\,keV rest-frame range probed here, taking into account the presence of a primary X-ray continuum with $\Gamma\approx 1.8$ and a ubiquitous soft X-ray excess (i.e. steepening) at energies below 1--2\,keV (e.g. \citealt{laor1997,brightman2013,ricci2017,trakhtenbrot2017}). In reality, the spectral slope may vary from one quasar to another. 
If we conservatively assume that individual $\Gamma$ values in our sample are distributed according to a normal distribution around $\Gamma=2$ with a standard deviation of 0.3, and take into account the eROSITA spectral response and the probed range of Galactic absorption, then this would result in an additional systematic uncertainty of less than 4\% (8\%) in the estimated absorption-corrected fluxes in the observed 0.3--2.3\,keV band for 90\% (99\%) of the sample.
This is much smaller than typical statistical uncertainties for our objects, which justifies our neglect of the uncertainty associated with X-ray spectral diversity. We also ignore any intrinsic X-ray absorption, since we are dealing with broad-line quasars only, and additionally exclude quasars with broad absorption lines (see Sect.~\ref{s:cleansample}).

We also imposed a limit on the optical flux, namely that magnitudes in the SDSS $g$-band $m_g<m_{g,{\rm max}}=19$. To this end, we used the $g$-band photometric measurements from the SDSS catalogue, namely point-spread-function (PSF) magnitudes, as recommended for point sources, such as the SDSS DR16Q quasars (by construction of this catalogue). This removes 7919 quasars (out of 12969). In addition, we excluded 20 quasars with incorrect magnitude measurements (negative values) from the SDSS DR16Q catalogue table from \cite{lyke2020}. Note that we could increase our sample by going deeper in the optical flux, at least down to $m_g\sim 19.5$, while retaining nearly the same high level of completeness (see \ref{appendix:A}). We decided not to do it, because the sample would then become strongly unbalanced in terms of X-ray and optical--UV depth, namely it would be heavily skewed towards one side of the X-ray--UV luminosity relation that we seek, which we would like to avoid.

For all quasars, we use the spectroscopic redshifts from \cite{wu2022spec} (ZSYS column in their catalogue), denoted hereafter as $z$. These authors have improved the redshift measurements reported in the original SDSS DR16Q catalogue by accurately accounting for velocity shifts of various lines with respect to the systemic velocity and correcting catastrophically wrong redshifts. Namely, the redshifts were corrected by more than 10000 km\,s$^{-1}$ for 1943 objects and by more than 1500 km\,s$^{-1}$ for $\approx 8000$ objects out of the 750414 quasars in the DR16Q catalogue. We also note that if two or more SDSS spectra are available for a given object, the SDSS DR16Q catalogue uses the best available spectrum, as defined by the SDSS pipeline \citep{Bolton_2012}.

We expect to have approximately four (less than seven with a probability of 90\%) spurious eROSITA--SDSS matches (upon application of the X-ray and optical flux limits). We obtained this estimate by assigning small, but much larger than $r98$, angular offsets (the result is virtually independent of the offset in the tested range from $1$\arcmin to $20$\arcmin) to the SDSS DR16Q quasars and then counting the number of cross-matches with the eROSITA catalogue.

Our choice of X-ray and optical flux thresholds has allowed us to obtain a large and statistically well defined sample, representative of both the bright and dim parts of the quasar luminosity function (LF). The resulting `raw' sample comprises 5050 quasars. 

It is worth noting that this quasar sample contains just a small ($\sim 15$\%) fraction of all AGN detected by eROSITA in the studied area of the sky above the (high) adopted X-ray flux limit (see also \citealt{2013A&A...558A..89K}). The remaining X-ray bright AGN have not passed our selection criteria, namely (i) do not qualify for the SDSS DR16Q catalogue (i.e. not a point-like source or no broad emission lines in the spectrum) or (ii) fall below our optical flux threshold. The majority of such objects are expected to be relatively nearby ($z\lesssim 0.5$) Seyfert I and Seyfert II galaxies.

\subsection{Cleaning the sample}
\label{s:cleansample}

\begin{table} 
	\vspace{2mm}
	\centering
	
	\vspace{2mm}
	\begin{tabular}{ m{5.5cm} m{2.0cm} } \hline
		{Catalogue}&  Number of counterparts \\ \hline
		{Roma-BZCAT Multifrequency Catalogue of Blazars}   &   147    \\
		{4th Fermi/LAT catalogue of AGN DR3}  &   50  \\
		{Blazar Radio and Optical Survey}      &    141   \\
		{Combined Radio All-Sky Targeted Eight GHz Survey} &    141\\ \hline
		{All of the above}      &    249  \\
		
        \hline
	\end{tabular}
    \caption{Cross-correlation with blazar catalogues.}
    \label{tab:blazars}
\end{table}
We applied some additional cleaning to the raw sample in order to improve its purity and statistical completeness. This included the following steps. 
\begin{itemize} 

\item
First, we excluded 161 X-ray sources that have multiple counterparts in the Gaia Data Release 3 (DR3) catalogue \citep{vallenari2023gaia} within $r98$ of the eROSITA position, because in such cases we cannot be sure that the X-ray emission is associated with the SDSS DR16Q quasar rather than with another nearby optical object.

\item
We cleaned the sample from blazars (BL Lac objects and flat-spectrum radio quasars). These are AGN with a highly collimated emission component that is associated with a relativistic jet directed towards the observer, while the current study is focused on the relationship between the emission from the accretion disc and its corona. We used the same approach as in our previous work \citep{prokhorenko2024}. Specifically, we cross-matched our sample with the 5th edition of the Roma-BZCAT Multifrequency Catalogue of Blazars \citep{massaro20155th}. This yielded 147 counterparts. We also cross-correlated our sample with the fourth catalogue of AGN detected by the Fermi Large Area Telescope, Data Release 3 \citep{ajello2022fourth}, namely with objects classified as `bll' (BL Lac-type objects), `bcu' (blazar candidates of unknown types) or `fsqr' (Flat Spectrum Radio Quasars). This resulted in another five counterparts. We then cross-matched our sample with the Blazar Radio and Optical Survey (BROS) \citep{itoh2020blazar}, excluding Gigahertz peaked-spectrum sources and compact steep-spectrum radio sources (based on the corresponding flag in the catalogue) from the search. This provided an additional 51 sources. Finally, we cross-correlated our sample with the Combined Radio All-Sky Targeted Eight GHz Survey \citep{healey2007crates}, which yielded another 46 counterparts. In all these cases, the search was done within 10\arcsec\ of the SDSS DR16Q optical positions to take into account that some of the positions given in the blazar catalogues are not very precise (in particular, when only a radio position is available). The expected total number of spurious cross-matches with the blazar catalogues is less than one with a probability of 90\%. In total, we found 249 blazars or blazar candidates associated with our objects (see Table~\ref{tab:blazars} for details) and removed them from the sample. 

\item
We excluded 72 broad-absorption-line (BAL) quasars, namely those that have a BAL probability greater than $0.2$ and BALnicity index for CIV greater than $0$, as quoted in \cite{lyke2020}. Such absorption lines indicate the presence of outflowing material, whose influence on the attenuation of the quasar's UV emission is difficult to assess. Note that information on the presence of BAL features is available only for quasars at $z>1.5$, for which the potential CIV absorption systems are within the wavelength range of the SDSS spectrometer. Hence, we do not perform any BAL cleaning for the quasars at $z<1.5$ in our sample.

\item
We leave only quasars at redshifts $0.5<z<2.5$ because their rest frame $2500$\,\AA\, falls into the 3800--9100\,\AA\ waveband of the SDSS-I/II spectrograph, which was used for spectroscopy of most quasars in our sample (the BOSS spectrograph, which was employed in SDSS-III and SDSS-IV, has a wider spectral range of 3650--10400\,\AA). This led to the removal of another 1453 quasars. Although we actually estimate the UV (2500\,\AA) luminosities of our quasars based on SDSS photometry rather than spectroscopy (see below), we nevertheless apply this redshift filter because we wish to account for the systematic uncertainty associated with the conversion of $g$-band photometric fluxes to UV luminosities (see Sect.~\ref{s:lums} below). 

\item 
Some sources in our sample may be multiply imaged lensed quasars or pairs and groups of physically connected or closely projected quasars. To clean the sample from such sources, we first excluded two sources with the Renormalised Unit Weight Error (ruwe) in Gaia DR3 catalogue larger than 1.3, which is a strong indication that the source is multiple or extended. Then, we correlated our sample with several catalogues of lensed, dual and binary quasars using a 2\arcsec\ radius \citep{DULAG,chen2025binaries,jing2025DESIldr1,makarov2022catalog,lemon2020lensedQ,lemon2023gl}, which led to the removal of another six sources. 

\end{itemize}

The above selections leave 3109 quasars in the sample. 

\subsection{Estimation of the X-ray and UV luminosities and construction of the final sample}
\label{s:lums}

\begin{figure}
    \begin{center}
        \vspace{\wid}
        \includegraphics[width=\columnwidth]{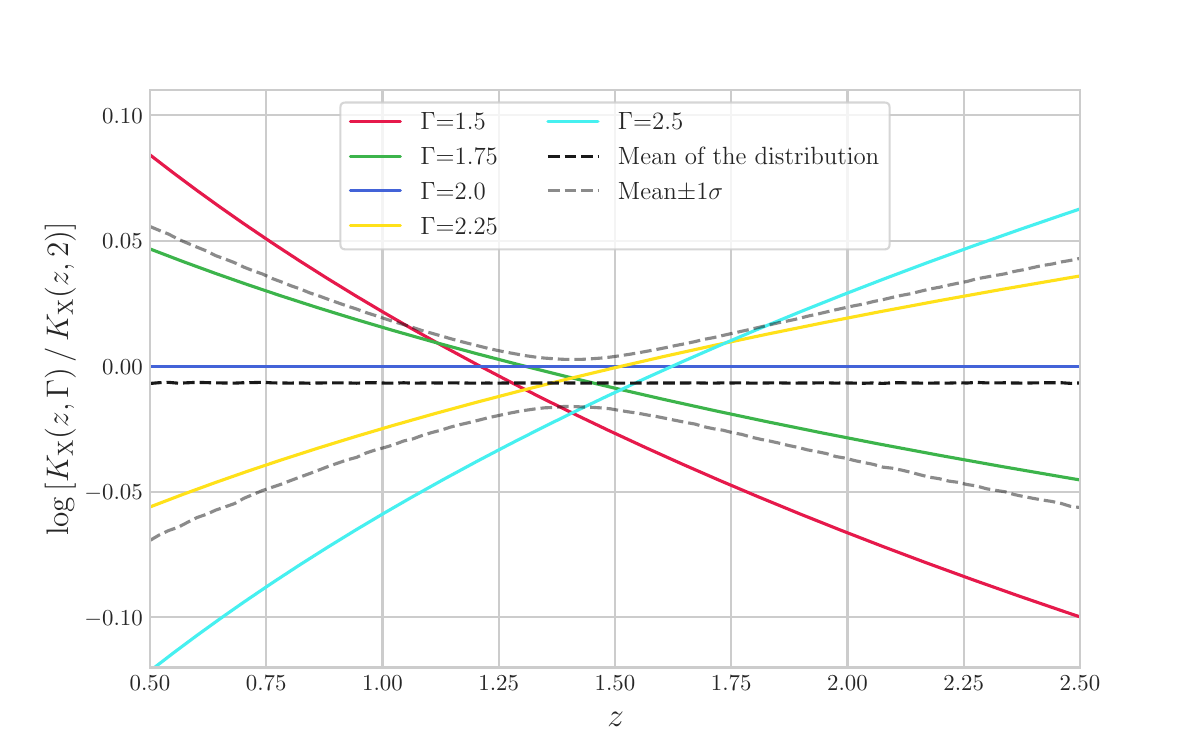}
        \caption{Solid lines show $\log{K_{\rm X}(z,\Gamma)/K_{\rm X}(z,2)}$ as a function of $z$ for different X-ray spectral indexes $\Gamma$, as indicated in the legend. The black dashed line marks the mean values of the $\log{\left[K_{\rm X}(z,\Gamma)/K_{\rm X}(z,2)\right]}$ distribution as a function of $z$ and assuming that $\Gamma$ follows a normal distribution with $\mu=2$ and $\sigma=0.3$. The black dashed semitransparent lines mark the mean plus one standard deviation and the mean minus one standard deviation as a function of $z$. Hereafter, $\log$ means $\log_{10}$. }
        \label{fig:KxGamma}
    \end{center} 
\end{figure}

As was already mentioned in Section~\ref{s:introduction}, we define $\Lx$ as the monochromatic luminosity (per frequency) at rest-frame 2\,keV, i.e. $\Lx\equiv L_\nu(2\,\textrm{keV})$.
It can be determined from the measured redshifts and unabsorbed X-ray fluxes in the observed 0.3--2.3\,keV band, using the corresponding $k$-correction. For a power-law spectrum with slope $\Gamma$ and redshift $z$, the latter is equal to
\begin{equation}
K_{\rm X}(z,\Gamma)=\frac{4.136\times10^{-18}\,\rm Hz^{-1}}{2^{\Gamma-1}(1+z)^{1-\Gamma}}\times\begin{cases}
\frac{1}{\ln(2.3/0.3)} &\rm if ~~\Gamma=2,\\
\frac{2-\Gamma}{2.3^{2-\Gamma}-0.3^{2-\Gamma}} & \rm otherwise.
\end{cases}
\end{equation}
 Figure~\ref{fig:KxGamma} shows $K_{\rm X}(z,\Gamma)$ for several slopes.

As was discussed in Section~\ref{s:rawsample}, the intrinsic X-ray spectral slope may vary from one object to another. This has a negligible systematic effect on our estimation of unabsorbed fluxes in the observed 0.3--2.3\,keV band from detector counts (see Sect.~\ref{s:rawsample}), but may significantly affect $K_{\rm X}(z,\Gamma)$ and hence our $\Lx$ estimates. To evaluate this effect, we assumed that the spectral slopes in our sample follow a normal distribution centered at $\Gamma=2$ with a standard deviation of 0.3, as we already did in Section~\ref{s:rawsample}. We simulated the $\log{\frac{K_{\rm X}(z,\Gamma)}{K_{\rm X}(z,2)}}$ distribution and calculated its mean value and standard deviation as a function of redshift (see Fig.~\ref{fig:KxGamma}). The mean equals $-0.0067$ (with the needed precision) independent of $z$, implying that the X-ray luminosity evaluated under our assumption of $\Gamma=2$ will be slightly biased with respect to the expected value of $\Lx$ for the adopted distribution of $\Gamma$. Hence, we define 
\begin{equation}
\label{eq:Lumx}
\Lx=0.985\,\frac{4\pi D_L^2(z) F_{{\rm X}}}{1+z}K_{\rm X}(z,2), 
\end{equation}
where $D_L(z)$ is the luminosity distance.

The fact that the systematic error on $\Fx$ is small allows us to estimate the total uncertainty of $\log\Lx$ as the square root of the sum of the squares of the statistical error of $\log\Fx$ (due to detector count fluctuations) and the systematic uncertainty of $\log{\frac{K_{\rm X}(z,\Gamma)}{K_{\rm X}(z,2)}}$. We define the latter as the standard deviation of the simulated distribution for each quasar's redshift. It is assumed here (and simple simulations we have conducted provide support for that) that the $\log\Lx$ distribution is close to a normal one even if $\log{\frac{K_{\rm X}(\Gamma)}{K_{\rm X}(2)}}$ is significantly skewed (as is the case for $z\in\left(1.1,1.5\right)$), because it convolves with the nearly normally distributed $\log\Fx$. As shown in Fig.~\ref{fig:KxGamma}, the systematic uncertainty in $\log\Lx$ is especially important (reaching $\approx 0.05$) for the nearest ($z\approx 0.5$) and most distant ($z\approx 2.5$) objects in our sample. 

We next define $\Luv$ as the monochromatic luminosity (per frequency) at rest-frame 2500\,\AA, i.e. $\Luv\equiv L_\nu(2500\,\textrm{\AA})$, and this quantity can be determined from measured distance and flux via 
\begin{equation}
\Luv=\frac{4\pi D_L^2(z)F_\nu(2500(1+z)\,\textrm{\AA})}{1+z}, 
\end{equation}
where in the numerator is the measured flux density per frequency at an observed wavelength of $2500(1+z)$\AA.

Importantly, we estimate $\Luv$ based on SDSS photometry rather than SDSS spectroscopy. This allows us to reliably control the statistical completeness of the sample. Indeed, quasar candidates for SDSS spectroscopy had been selected based on their photometric magnitudes and colours and we know (see \ref{appendix:A}) how the completeness of this selection depends on the apparent magnitude limit in the $g$ band. Therefore, since we estimate $\Luv$ at the epoch of target selection (SDSS photometry), we can specify a strict UV luminosity limit as a function of redshift, $L_{\rm 2500min}(z)$, above which our sample is statistically complete. This would be impossible to do if we instead used SDSS spectroscopy for estimating $\Luv$, since SDSS spectra were taken several years after the corresponding photometric observations, when the quasars had already changed their luminosity due to intrinsic variability. 

Specifically, we estimated $F_\nu(2500(1+z)\,\textrm{\AA})$, for each quasar based on its SDSS $g$-band photometric measurement ($m_{g}$), SDSS spectroscopic redshift ($z$) and the corresponding $k$-correction, $K_{\rm UV}(z)$:
\begin{equation}
F_\nu(2500(1+z)\,\textrm{\AA})=F_{\rm mg}\,K_{\rm UV}(z),
\label{eq:k}
\end{equation}
where $F_{\rm mg}$ is the flux corresponding to $m_{g}$.

Assuming that the spectral shape did not change between the epochs of SDSS photometry and spectroscopy for each quasar, we estimated the $K_{\rm UV}$ values using the aforementioned catalogue of spectral properties by \cite{wu2022spec}, which is based on SDSS spectroscopy. These authors used pyQSOFIT (see \citealt{Shen_2019}) and fitted the SDSS spectra by a global continuum plus emission lines model. For the continuum they used a sum of a power-law, a third-order polynomial model and a component responsible for optical and UV FeII emission. 

We used the parameters of the power-law and third-order polynomial quasi-continuum for each quasar to estimate 
$F^{\rm spec}_\nu(2500(1+z)\,\textrm{\AA})$ (hereafter, the superscript `spec' indicates that fluxes are estimated from spectra rather than from photometry) for the continuum emission, excluding the contribution of Fe II. Then we calculated $K_{\rm UV}$ as the ratio of the 
$F^{\rm spec}_\nu(2500(1+z)\,\textrm{\AA})$ and the integrated flux (including the contribution of emission lines and FeII emission) over the observed $g$ band, $F^{\rm spec}_g$: 
\begin{equation}
K_{\rm UV}=\frac{F^{\rm spec}_\nu(2500(1+z)\,\textrm{\AA})}{F^{\rm spec}_{g}}.
\label{eq:kspec}
\end{equation}

We use in our analysis the continuum rather than the total luminosity at $2500$\,\AA, since the continuum emission presumably comes from the innermost region of the accretion disc near the corona and it varies roughly in concert with the coronal X-ray emission on the time scales we are interested in (see Sect.~\ref{s: Basic considerations}). However, the difference between the two quantities is modest: namely, the total flux density at rest-frame $2500$\,\AA\ is greater than 
$F^{\rm spec}_\nu(2500(1+z)\,\textrm{\AA})$ by just $15$\% on average and this fraction ranges from 8\% to 27\% for 90\% for our quasars. We also note that the contribution of emission lines and FeII emission to the total flux in the $g$ band is 13\% on average and ranges from 6\% to 27\% for 90\% of our sample. We neglect any intrinsic extinction in the UV, assuming that it is small for our quasars since they all have broad optical emission lines. It is thus clear that we underestimate the intrinsic UV luminosities, but the amplitude of this effect is difficult to assess. Note that a similar small systematic bias (due to intrinsic absorption) should affect our estimates of the X-ray luminosities, as was already mentioned in Section~\ref{s:rawsample}.  

We then used the 
$F_\nu(2500(1+z)\,\textrm{\AA})$ values to estimate $\Luv$. 
The uncertainty of $\Luv$ is composed of the photometry uncertainty $\sigma(m_{g})$ and the uncertainty in the $k$-correction, $\sigma(\log{K_{\rm UV}})$. The latter is derived from the uncertainties in 
$F^{\rm spec}_\nu(2500(1+z)\,\textrm{\AA})$ and $F^{\rm spec}_g$ by propagation of error in equation~(\ref{eq:kspec}), neglecting the uncertainty in spectroscopic redshift. We adopted the 
$F^{\rm spec}_\nu(2500(1+z)\,\textrm{\AA})$ uncertainty directly from the relevant LOGL2500\_ERR column in \cite{wu2022spec}, and estimated the $F^{\rm spec}_g$ uncertainty by averaging the uncertainty in the spectral bins across the $g$-band. Hence, 
\begin{equation}
\sigma\left(\log{F_\nu(2500(1+z)\,\textrm{\AA})}\right)=\sqrt{\left[0.4\sigma(m_{g})\right]^2+\sigma(\log{K_{\rm UV}})^2}.
\label{eq:sigluv}
\end{equation}
During this procedure, we removed from our sample 50 quasars with missing LOGL2500\_ERR values, which indicates that the spectrum breaks off around 2500\,\AA, so the uncertainty cannot be evaluated. It turns out that the flux uncertainty 
$\sigma\left(\log{F^{\rm spec}_\nu(2500(1+z)\,\textrm{\AA})}\right)$
is dominated by photometry. The median value of $0.4\sigma( m_{g})/\sigma(\log{K_{\rm UV}})$ is 3.5.

\begin{figure}[t]
    \begin{center}
        \vspace{\wid}
        \includegraphics[width=\columnwidth]{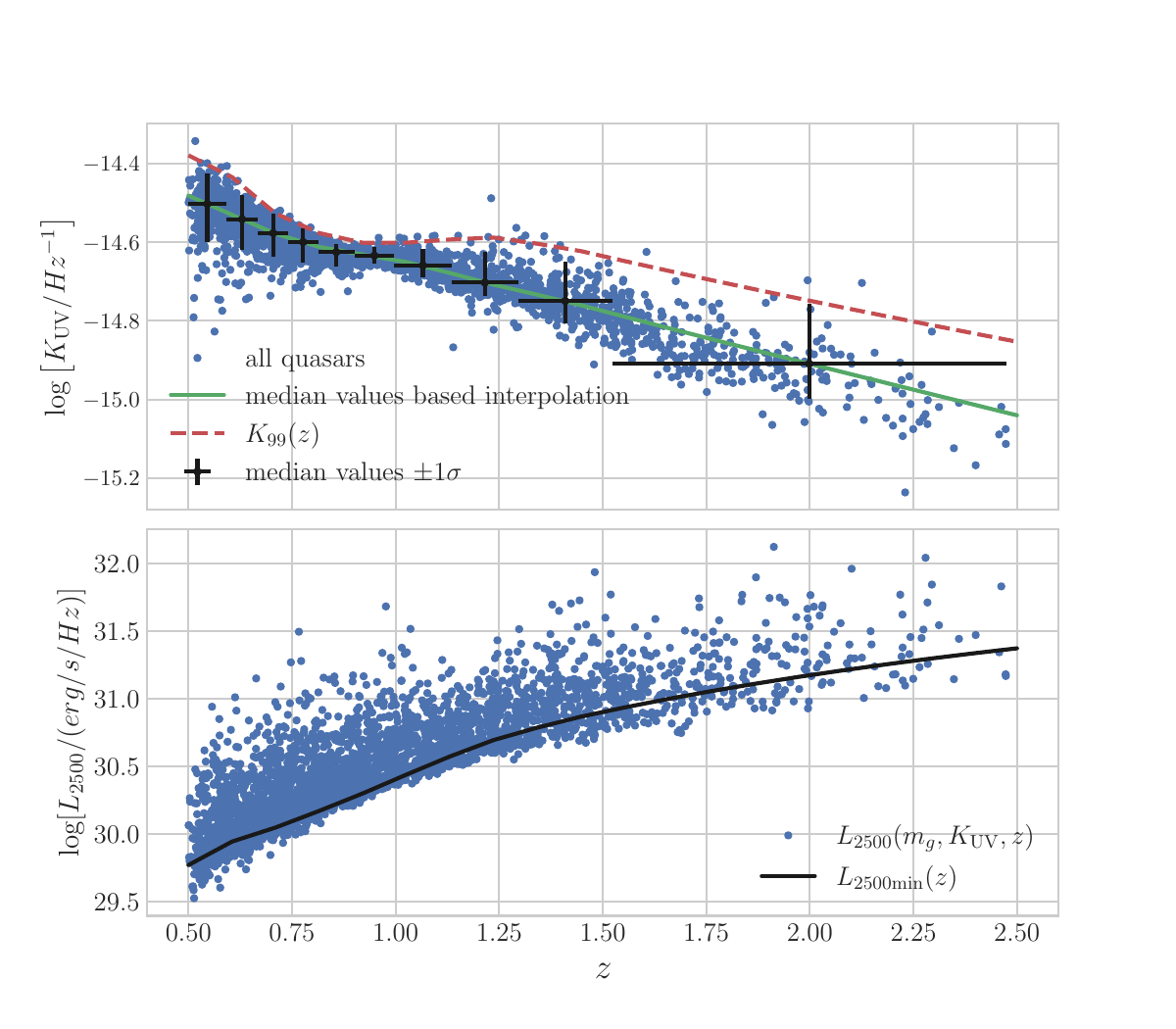}
        \caption{Top panel: Scatter-plot of $k$-correction vs. redshift for quasars from our sample before the additional UV luminosity selection. The red dashed line bounds the region containing 99\% of objects in each redshift bin. Bottom panel: Corresponding scatter-plot of 2500\AA\ luminosity vs. redshift. The black line depicts the luminosity limit $L_{\rm 2500min}$ as a function of $z$.}
        \label{fig:sampleLmin}
    \end{center} 
\end{figure}

We finally intend to impose a strict luminosity limit $L_{\rm 2500min}(z)$ on our sample, in order to minimize the impact of any $k$-correction (i.e. optical--UV spectral shape) related selection bias on our analysis of the X-ray--UV luminosity correlation. 
To this end, we divided our quasar sample into ten redshift bins containing nearly the same number of objects (from 313 to 316 objects per bin). As is shown in the top panel of Fig.~\ref{fig:sampleLmin}, the scatter in the $k$-correction within a given redshift bin is not large. Namely, 90\% (5--95\%) of quasars have $k$-corrections differing by less than $0.25$ (less than $0.18$ for all bins except the highest redshift one). For each bin, we determined a limiting value $K_{\rm 99}$ such that 99\% of quasars in that bin have $K_{\rm UV}<K_{\rm 99}$\footnote{This threshold is sufficiently conservative for our purposes as it is comparable to the aforementioned completeness ($\gtrsim 98$\%) of SDSS spectroscopic coverage of quasars.}. We then derived a smooth dependence $K_{\rm 99}(z)$ by interpolation between these discrete $K_{\rm 99}$ values (see the top panel of Fig.~\ref{fig:sampleLmin}). Finally, we defined a UV luminosity limit,
\begin{equation}
L_{\rm 2500min}(z)=\frac{4\pi \,D_L^2(z)\, F_{\rm mg=19} \,K_{\rm 99}(z)}{1+z},
\label{eq:luvmin}
\end{equation}
where $F_{\rm mg=19}$ is the flux in the $g$ band corresponding to $m_{g,{\rm max}}=19$, and excluded all sources (645) with $\Luv<L_{\rm 2500min}(z)$ from our sample (see the bottom panel of Fig.~\ref{fig:sampleLmin}). This ensures that quasars with $\Luv>L_{\rm 2500min}(z)$, selected for the subsequent analysis, will fairly represent the overall quasar population with all its spectral diversity. 

The resulting sample consists of 2414 quasars, and it is used in our analysis below. We will call it the \srg/eROSITA--SDSS quasar sample. 

\subsection{Properties of the quasar sample}
\label{s:properties}

\begin{figure}[h]
    \begin{center}
        \vspace{\wid}
        \includegraphics[width=\columnwidth]{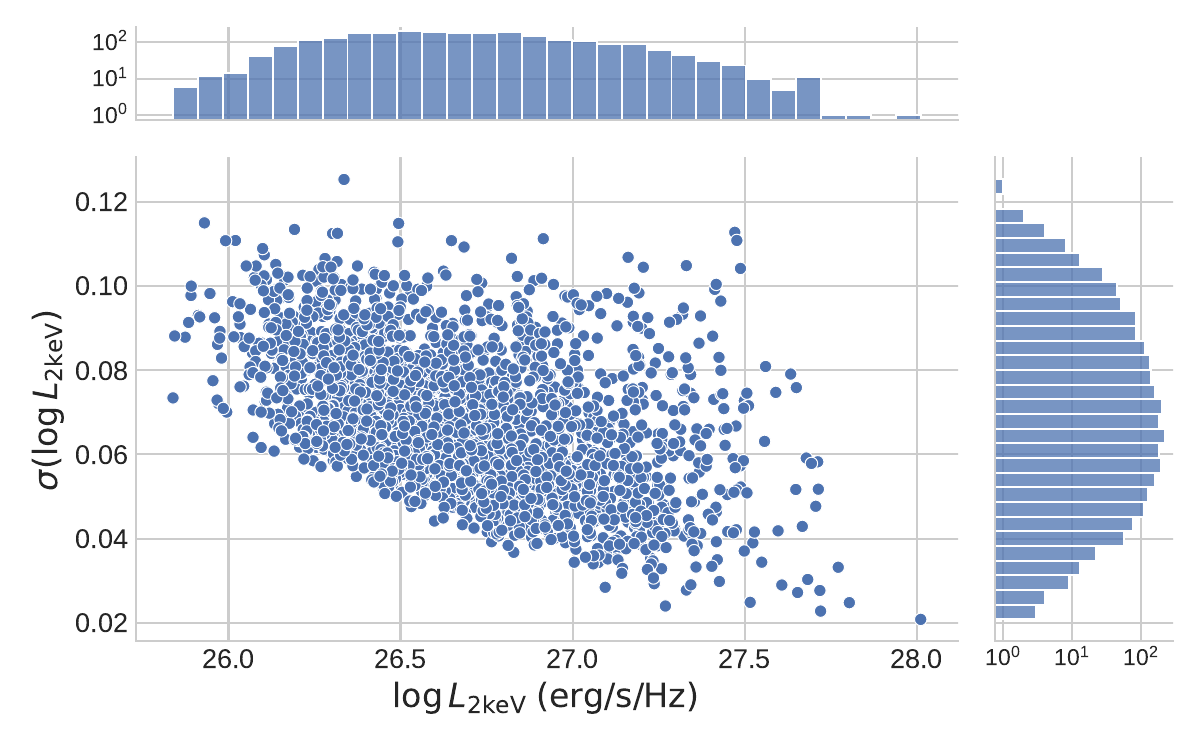}
        \caption{Scatter-plot of the logarithms of X-ray luminosities vs. their uncertainties for the \srg/eROSITA--SDSS quasar sample. The histograms shown on the sides reflect projections of the two-dimensional distribution on the corresponding axes (note the logarithmic scale for the number of objects).}
        \label{fig:Lxdistrs}
    \end{center} 
\end{figure}

\begin{figure}[t]
    \begin{center}
        \vspace{\wid}
        \includegraphics[width=\columnwidth]{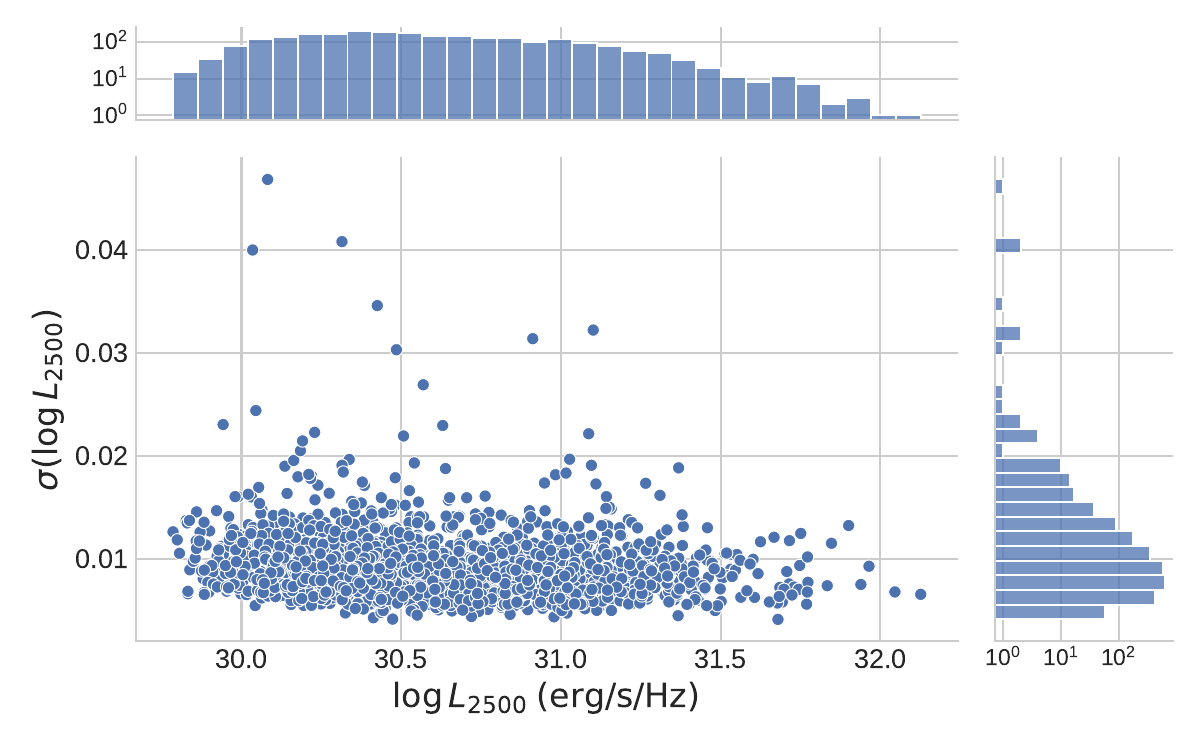}
        \caption{Scatter-plot of the logarithms of UV (2500\,\AA) luminosities vs. their uncertainties for the \srg/eROSITA--SDSS quasar sample. The histograms shown on the sides reflect projections of the two-dimensional distribution on the corresponding axes (note the logarithmic scale for the number of objects).}
        \label{fig:L2500distrs}
    \end{center} 
\end{figure}

Figure~\ref{fig:Lxdistrs} shows the distribution of the logarithms of the X-ray luminosities and the corresponding uncertainties, $\sigma\left(\log \Lx\right)$ (due to photon count fluctuations and the systematic uncertainty in the X-ray spectral shape), based on \srg/eROSITA data. The median value of $\left[\sigma\left(\log \Lx\right)\right]^2=4.5\times 10^{-3}$. Note that it would be $3.0\times 10^{-3}$ if we neglected the systematic uncertainty. Figure~\ref{fig:L2500distrs} shows the distribution of the logarithms of the UV luminosities and their uncertainties, $\sigma\left(\log{\Luv}\right)$. The latter is equal to the uncertainty in $\log{F_{2500}}$, defined in equation~(\ref{eq:sigluv}). The median $\left[\sigma(\log{\Luv})\right]^2=7.4\times10^{-5}$, much smaller than for the X-ray luminosity.
Note that we neglect uncertainties in the spectroscopic $z$ measurements and in the cosmological model.

\begin{figure}[h]
    \begin{center}
    	\vspace{\wid}
    	\includegraphics[width=1.1\columnwidth]{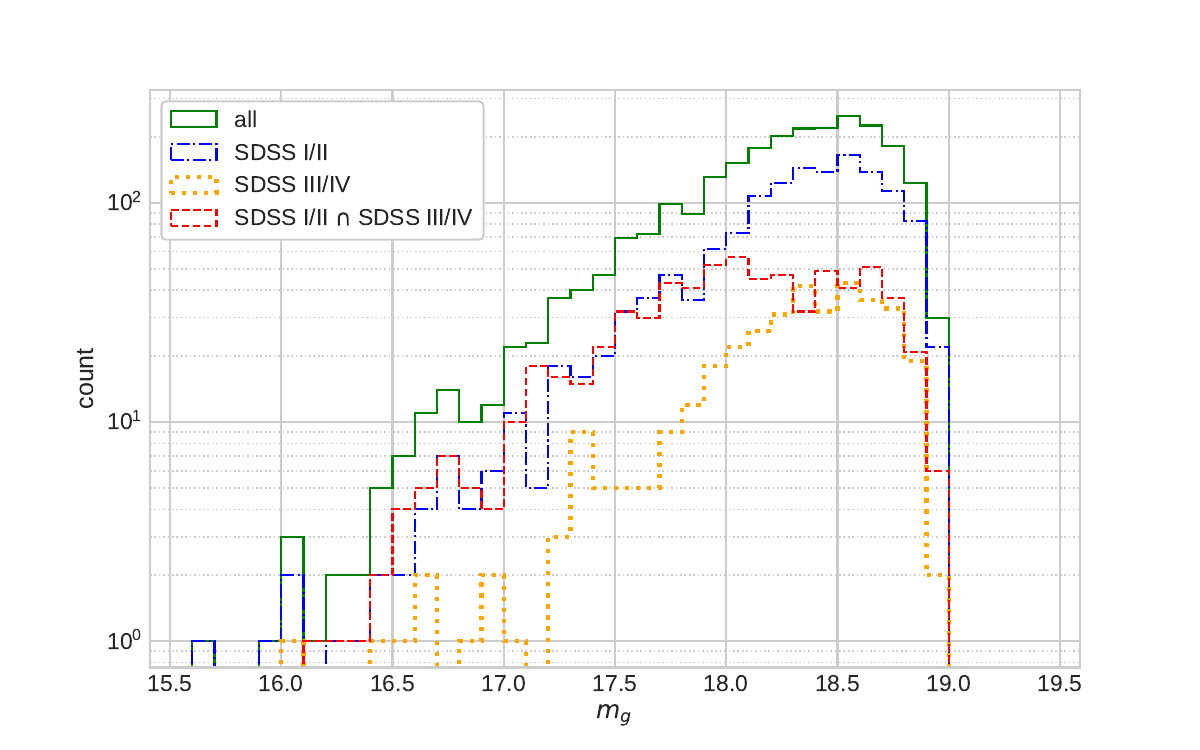}
    	\caption{Distribution of SDSS $g$-band photometric magnitudes for the \srg/eROSITA--SDSS quasar sample. The full sample is shown in green. The other histograms show subsamples according to SDSS spectroscopy: blue, quasars observed in SDSS I/II only; orange, those observed in SDSS III/IV only; red, those observed both in SDSS I/II and SDSS III/IV.
}
        \label{fig:mdistrs}
    \end{center}
\end{figure}

Figure~\ref{fig:mdistrs} shows the $g$-band photometric magnitudes measured by SDSS. We have checked that the apparent drop of the $m_g$ distribution near our adopted optical limit of $m_{g,{\rm max}}=19$ is not due to paucity of such quasars in the DR16Q catalogue but rather a consequence of the limits that we imposed on UV luminosity, $L_{\rm 2500min}$, and X-ray flux, $F_{{\rm X,min}}$ (via the intrinsic correlation between X-ray and UV luminosity of quasars). 

\begin{figure*}
    \begin{center}
    	\vspace{\wid}
    	\includegraphics[width=0.9\textwidth]{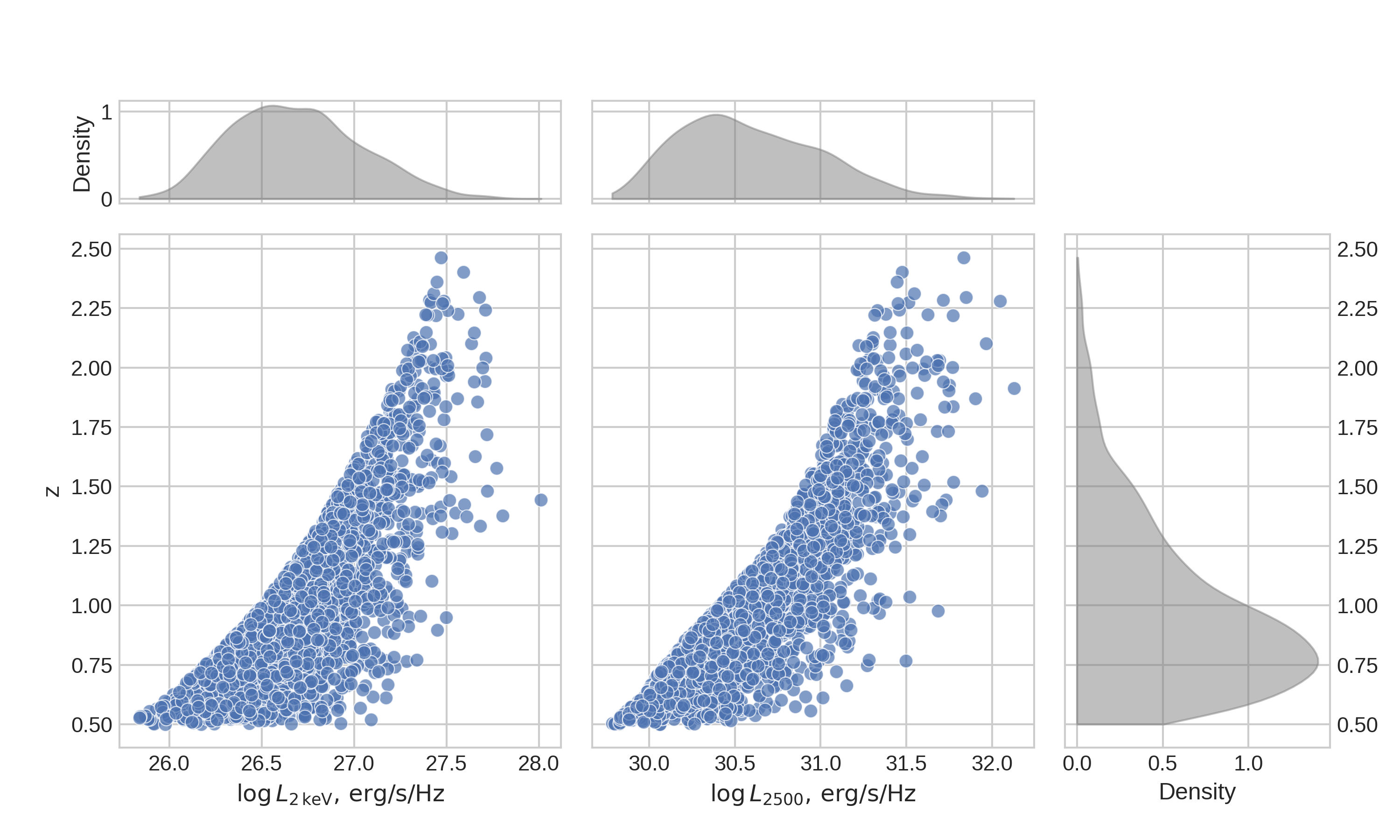}
    	\caption{Scatter-plots of redshift vs. X-ray luminosity (left) and UV luminosity (right), with the corresponding marginal distributions for the \srg/eROSITA--SDSS sample. }
        \label{fig:sampleLz} 
    \end{center}
\end{figure*}

Figure~\ref{fig:sampleLz} shows the distributions of: (i) redshifts, (ii) X-ray luminosities and (iii) UV luminosities.
The median redshift of our sample is 0.897. Importantly, we cover a broad dynamic range, $\sim 2$\,dex, in both X-ray and UV luminosity. 
The smooth boundary of the distribution in the middle panel reflects the fact that by defining $\Luv$ as a function of $m_g$ and $z$ we have achieved a strict cut in $\Luv$ for each $z$, just as for $\Lx$, which is important for the following statistical analysis.

\section{Method} 
\label{s:method}

The inferred parameters of the $\Lx$--$\Luv$ correlation can be affected by selection effects and the specifics of methods applied to the data. 
Although such biases have been addressed before, we see some potentially significant shortcomings in the previous analyses. 

In particular, \cite{dainotti2022} took into account flux selection in treating the cosmological evolution of quasars but not in determining the $\Lx$--$\Luv$ correlation parameters (based on de-evolved luminosities). Other authors attempted to take flux selection into account by considering quasars undetected in X-rays. In particular, \cite{lusso2016tight} and \cite{rankine2024} 
took into account X-ray flux upper limits for optically selected quasars, but did not discuss how the correlation might be affected by X-ray bright quasars that are undetected in the optical. 

We propose a novel method, which uses quasars detected both in the X-ray and optical--UV bands and fully accounts for biases associated with objects undetected in each of these bands. A similar approach was used, in a different context, by \cite{takeuchi2013}, who constructed a far-ultraviolet and far-infrared bivariate luminosity function of galaxies using a copula approach, which accounts for selection effects in a similar way to ours (see equation 25 in that article). 

Another novelty of our method consists in a more general definition of the scatter of the correlation between X-ray and UV luminosities compared to previous works. All preceding studies (e.g., \citealt{lusso2010,bisogni2021chandra,dainotti2022,signorini2024quasars,rankine2024}) attributed this scatter to either $\Lx$ or $\Luv$, i.e. to only one of the two physical quantities involved in the correlation. In doing so, it is assumed that either $\Lx$ depends on $\Luv$ or vice versa, and results will depend on this assumption. For example, in the ordinary least squares (OLS) analysis conducted by \cite{steffen2006x}, significantly different results were obtained when the scatter was ascribed to X-ray or to UV luminosity: the authors found slopes of $0.642\pm0.021$ and $0.815\pm0.023$ for OLS(X|UV) and OLS(UV|X), respectively. However, there is no obvious reason to regard either X-ray or UV luminosity as the dependent variable. Thus, we do not give any preference to either $\Lx$ and $\Luv$ and take the scatter in both $\Lx$ and $\Luv$ into account.

\subsection{Mathematical formulation}
\label{s:formulation}

For convenience and to facilitate comparison with previous studies, we will work in logarithmic luminosity space and 
thus introduce $\lx=\log\frac{\Lx}{\lumdU}$, $\luv=\log\frac{\Luv}{\lumdU}$.
Suppose that these `observed' luminosities are related to each other with some scatter. Then we can introduce `primary' luminosities  $\plx=\log\frac{\pLx}{\lumdU}$ and $\pluv=\log\frac{\pLuv}{\lumdU}$ such that one is related to the other via a continuous dependence, and $\lx$ and $\luv$ are scattered around this `primary' relation between $\plx$ and $\pluv$. The scatter arises due to (i) `intrinsic scatter', (ii) variability and (iii) flux measurement uncertainties. Therefore, by `intrinsic' we mean all sources of scatter that are not related to variability and flux measurement uncertainties. Specifically, we seek a correlation between X-ray and UV luminosities measured simultaneously and averaged over a certain time scale. The exact meaning of `simultaneity' and the `averaging time scale' will be revealed in Section~\ref{s:variability}. 

We further assume, for simplicity and computational efficiency, that all three sources of scatter, both for X-ray and UV luminosity, are mutually independent and normally distributed, that is the observed X-ray and UV luminosities obey lognormal distributions relative to the primary ones. Then the distribution of $ \lx$ for a given $\plx$ is a normal one: 
\begin{equation}
\mathcal{N}(\lx|\plx,\s{X})=\frac{1}{\sqrt{2\pi}\s{X}}\exp{\left(-\frac{\left(\plx-\lx\right)^2}{2\Sigx}\right)},
\label{eq:Vx}
\end{equation} 
and, similarly, $ \luv$ for a given $\pluv$:
\begin{equation}
\mathcal{N}(\luv| \pluv,\s{UV})=\frac{1}{\sqrt{2\pi}\s{UV}}\exp{\left(-\frac{\left(\pluv-\luv\right)^2}{2\Siguv}\right)},
\label{eq:Vuv}
\end{equation} 
where
\begin{align}
&\Sigx=\Sigmx+\Sigvarx+\Sigintx,\nonumber\\
&\Siguv=\Sigmuv+\Sigvaruv+\Sigintuv.
\label{eq:sigmas}
\end{align}
 Here $\s{mX}$ and $\s{mUV}$ are the X-ray and UV measurement uncertainties, $\s{varX}$ and $\s{varUV}$ the contributions associated with X-ray and UV variability, and $\s{intX}$ and $\s{intUV}$ account for the intrinsic scatter.

It should be emphasized that we do not know the nature of the intrinsic scatter and thus assume, for simplicity, that its contributions along the X-ray and UV luminosity axes are independent of each other. The real situation may be more complicated, with some cross-talk between X-ray and UV intrinsic scatter. Nevertheless, the current formulation is more general than if we had assumed the intrinsic scatter to be associated with either X-ray or UV luminosity. The contributions of X-ray and UV measurement  uncertainties obviously do not correlate with each other or with other contributions to the total scatter. The exact meaning of the X-ray and UV variability contributions to the scatter will be clarified in 
Section \ref{s:variability}.

Following previous studies (e.g. \citealt{lusso2016tight}), we seek the primary dependence in a linear form (in logarithmic space):
\begin{equation}
\label{eq:lxluv}
\plx = \beta +\gamma \pluv,
\end{equation}
where we assume that $\gamma$ and $\beta$ are independent of redshift (at least in the interval $0.5<z<2.5$). 

We also define a `primary quasar luminosity function (LF)':
\begin{equation}
\label{eq:TLF}
\pLF(\plx,z)\equiv\frac{dN(z)}{d\plx},
\end{equation} 
which is the number of quasars per comoving volume of the Universe per $\plx$ as a function of $z$, and correspondingly the normalized primary LF: 
\begin{equation}
\label{eq:plf}
\npLF(\plx|z)\equiv\frac{\pLF(\plx,z)}{\int\pLF(\plx,z)d\plx}.
\end{equation}
Here, the integration is to be carried out over $\plx$ from $-\infty$ to $\infty$. However, for computational convenience we set a lower limit on luminosity, $l_{\rm cut}$, below which there are no quasars in the Universe. That is, we additionally postulate that $\npLF(\plx|z)=0$ for $\plx<l_{\rm cut}$. The specific choice of this threshold does not affect our results if $l_{\rm cut}$ is less than the luminosity corresponding to the X-ray and optical/UV flux limits at a given redshift.

With these definitions, the probability density for a quasar at redshift $z$ to have observed luminosities $\lx$ and $\luv$ (without yet taking selection by luminosity thresholds into account) is
\begin{align}
\label{eq:fi}
f(\lx, \luv|z,\s{UV},\s{X}) =\int\int \mathcal{N}(\lx| \plx,\s{X}) \mathcal{N}(\luv| \pluv,\s{UV}) \nonumber\\ \times \delta\left(\plx -(\beta +\gamma \pluv)\right)|\gamma| \npLF(\plx|z) \, d\plx d \pluv \nonumber\\= \int \mathcal{N}(\lx|\plx,\s{X})\, \mathcal{N}(\luv|(\plx-\beta)/\gamma,\s{UV})\, \npLF(\plx|z) \, d\plx,
\end{align}
where $\delta$ is the Dirac delta function and the integration is again carried out from $-\infty$ to $\infty$.

Alternatively, we could consider the UV (at 2500\,\AA) LF, $\pLFuv(\pluv,z)$. 
The model given by equation~(\ref{eq:fi}) has the following meaning: the delta function determines the primary dependence, $\mathcal{N}$ describes the scatter around it along the X-ray and UV luminosity axes, and $\npLF$ describes the changing (normalized) number density of quasars along the primary dependence. 

\subsubsection{Account for selection effects}

Next, we wish to take into account the selection effects associated with the X-ray and UV luminosity thresholds. Suppose that: (i) $P(A_i)$ is the probability of having a quasar with X-ray luminosity $l_{\rm X,i}$ and UV luminosity $l_{\rm UV,i}$ in the Universe at redshift z, and (ii) $P(B)$ is the probability to pass the selection criterion. Then, the probability that a quasar with $l_{\rm X,i}$ and $l_{\rm UV,i}$ will enter the resulting sample is $P(B|A_i)$. When analyzing a sample of thus selected quasars, we are actually interested in another probability, $P(A_i|B)$, which can be found using Bayes' theorem:
\begin{equation}
\label{eq:Bayess}
P(A_i|B)=\frac{P(B|A_i)P(A_i)}{P(B)}=\frac{P(B|A_i)P(A_i)}{\sum_j P(B|A_j)P(A_j)}.
\end{equation} 
Here, the summation in the denominator on the right-hand side is carried out over all possible pairs of $l_{\rm X,j}$ and $l_{\rm UV,j}$. 

We can rewrite this equation in terms of probability density by replacing $P(A_i)$ and $P(B|A_i)$ by $f(l_{\rm X,i}, l_{\rm UV,i}|z_i,\s{UV,i},\s{X,i})d\luv d\lx$ and $\rho(l_{\rm X,i}, l_{\rm UV,i},z_i)$, respectively, where $\rho$ is a `selection function' for given $l_{\rm X,i}$, $l_{\rm UV,i}$ and $z_i$. Then the observed probability density $f_{\rm obs}$, i.e. the probability density of finding a quasar with particular $l_{\rm UV,i}$ and $l_{\rm X,i}$ at given $z_i$, provided it has been selected (based on its $l_{\rm X,i}$, $l_{\rm UV,i}$ and $z_i$), is:
\begin{align}
\label{eq:fobs}
f_{{\rm obs},i}\equiv f_{{\rm obs}}\left(l_{\rm X,i}, l_{\rm UV,i}|z_i, \sigma^2_{{\rm mX,i}}, \sigma^2_{{\rm mUV,i}}, \sigma^2_{{\rm varX,i}},\sigma^2_{{\rm varUV,i}}\right)\nonumber \\=\frac{\rho(l_{\rm X,i}, l_{\rm UV,i},z_i)f(l_{\rm X,i}, l_{\rm UV,i}|z_i,\s{UV,i},\s{X,i})}{\int \int \rho(l_{\rm X}, l_{\rm UV},z_i)f(\lx, \luv|z_i,\s{UV,i},\s{X,i})d\luv d\lx},
\end{align} 
where, in our case, $\rho(l_{\rm X}, l_{\rm UV},z)=\Theta\left(l_{\rm X}-l_{\rm Xmin}(z)\right)\Theta\left(l_{\rm UV}-l_{\rm UVmin}\left(z\right)\right)$ is a two-dimensional step function. In fact, there should be a multiplier responsible for the $z$ limits we applied ($0.5<z<2.5$), but it does not depend on the luminosities, so it cancels out in the numerator and denominator of equation~(\ref{eq:fobs}). 

In our case, the limiting X-ray luminosity at $z_i$ is determined by the \srg/eROSITA flux limit as $l_{{\rm Xmin}}(z_i)=\log{\frac{0.985\,4\pi D_L^2(z_i) F_{{\rm Xmin}}K_{\rm X}(z_i)}{(1+z_i)(\lumU)}}$ [in accordance with equation~(\ref{eq:Lumx})], where $K_{\rm X}(z)$ is the $k$-correction from the observed-frame 0.3--2.3\,keV flux to the flux density per frequency at observed $2/(1+z)$\,keV (using our adopted power-law model with photon index $\Gamma=2$). The limiting UV luminosity $l_{\rm UVmin}(z_i)\equiv L_{\rm 2500min}(z_i)$ was specified in Section~\ref{s:lums}.

As a result, equation~(\ref{eq:fobs}) simplifies to:
\begin{align}
\label{eq:fobs2}
f_{{\rm obs},i}
=\frac{ f(l_{{\rm X},i}, l_{{\rm UV},i}|z_i,\s{UV,i},\s{X,i})}{\int_{l_{\rm UVmin}(z_i)}^{\infty} \int_{l_{\rm Xmin}(z_i)}^{\infty} f(\lx, \luv|z_i,\s{UV,i},\s{X,i})d\lx d\luv},
\end{align} 
where we have omitted the theta functions in the numerator because we are sampling quasars that by definition are above the imposed luminosity thresholds. Thus, the observed probability density $f_{\rm obs,i}$ for quasar $i$ from our sample is $f(l_{{\rm X},i}, l_{{\rm UV},i}|z_i,\s{UV,i},\s{X,i})$ normalized by the integral over the $[\luv,\lx]$ parameter area above the luminosity thresholds at a given $z_i$ (see also \citealt{takeuchi2013}).

Crucially for the purposes of this study, the integration in equation~(\ref{eq:fi}) properly accounts for flux limit related selection biases. Objects with given $\plx$ or $\pluv$ can have higher or lower observed $\lx$ or $\luv$ due to (i) intrinsic scatter around the primary dependence between $\plx$ or $\pluv$, (ii) variability and (iii) photon count fluctuations at the detector (the latter, combined with the higher space density of less luminous objects leads to the well-known Eddington bias). In our method, we self-consistently account for all these effects.

Furthermore, the absence of a redshift integral in equation~(\ref{eq:fobs}) renders our method insensitive to redshift selection. 
Essentially, we adjust each quasar to the probability density corresponding to its $z_i$, and the relative number density of quasars at different redshifts does not play a role. Thus, the lack of objects at some redshift in the sample would merely lead to less information about quasars at this $z$, so that such objects would contribute less statistical weight to the final result. Therefore, we could also, for example, split our sample into bins by $z$, and apply our method to each bin separately, without applying a redshift selection to the algorithm. We finally note that the SDSS quasar catalogue is actually not significantly affected by redshift selection in our adopted parameter range of $m_g<19$ and $0.5<z<2.5$ (see \ref{appendix:A}).

\subsubsection{Final algorithm}
\label{s:finalalg}

Eventually, we  use the following likelihood function for the \srg/eROSITA--SDSS quasar sample:
 
\begin{align}
\label{eq:likelihood}
&\mathcal{L}\left(\beta,\gamma, \Sigintuv, \Sigintx,{\pmb p_{LF}}\right)=
\prod_if_{{\rm obs}}(l_{\rm X,i}, l_{\rm UV,i}|z_i, \sigma^2_{{\rm mX,i}}, \sigma^2_{{\rm mUV,i}},& \nonumber \\&\sigma^2_{{\rm varX,i}},\sigma^2_{{\rm varUV,i}};\beta,\gamma, \Sigintuv, \Sigintx,{\pmb p_{LF}}),&
\end{align}
where the summation is carried out over all quasars in the \srg/eROSITA--SDSS sample 
, $\pmb p_{LF}$ are parameters of the LF, which we define later.

In practice, we could fix the quasar $\pLF$ to some model from the literature and consider only the four parameters of the correlation, namely, $\beta$, $\gamma$, $\Sigintuv$ and $\Sigintx$ as free ones. Alternatively, we could also simultaneously derive the parameters of $\pLF$. We adopted the latter approach and used a Markov chain Monte Carlo (MCMC) method for Bayesian inference of parameter distribution, unless otherwise specified. Specifically, we utilize the Python package {\it emcee} \citep{foreman2013emcee}, an implementation of the affine-invariant ensemble sampler for MCMC proposed by Goodman and Weare. 

\subsection{Luminosity function}
\label{s:luminosityfunction}

Our method relies on assumptions about the AGN LF, which limits the precision of determination of the X-ray--UV luminosity correlation parameters. On the other hand, this allows us to test various hypotheses about the LF based on the available \srg/eROSITA and SDSS data, i.e. we can control the impact of uncertainties in the LF on the inferred correlation parameters. 

As introduced in Section~\ref{s:formulation}, we based our analysis on the primary LF, $\pLF$, which gives the number density of quasars per $\plx$. However, observational studies always report on LFs that are defined as number density per observed X-ray (or optical/UV) luminosity, $\lx$. This observed X-ray luminosity function, $LF$, is different from our $\pLF$ due to the intrinsic scatter of the X-ray--UV luminosity correlation and due to the broadening caused by variability. Namely, $NLF$ (normalized $LF$) results from convolution of $\npLF$ with a Gaussian of width $\sqrt{\Sigintx+\Sigvarx}$:
\begin{align}
\label{eq:fdd}
NLF(\lx|z) = 
 \int \mathcal{N}(\lx|\plx,\sqrt{\Sigintx+\Sigvarx}) \npLF(\plx|z) \, d\plx,
\end{align} 
where $\mathcal{N}(\lx| \plx,\sqrt{\Sigintx+\Sigvarx})$ is the normal distribution defined in equation~(\ref{eq:Vx}). We can thus expect $NLF$ to be different from $\npLF$, especially in the (steep) high-luminosity part. This is demonstrated explicitly in \ref{appendix:B}. 

In what follows, we use a luminosity-and-density-evolution (LADE) model, following \citet{aird2015}, as our baseline $\pLF$ model (in the X-ray band). Since, as explained above, we are not sensitive to the redshift evolution of quasar number density, we cannot limit the parameters describing the evolution of the LF normalization.
Thus, we parameterize the model as follows:
\begin{equation}
\label{eq:LumF}
\pLF(\plx,z)\propto \left[ \left( \frac{\pLx}{e(z) L^*}\right)^{g_1}+\left(\frac{\pLx}{e(z) L^*}\right)^{g_2}\right]^{-1},
\end{equation}
where $g_1$ and $g_2$ are the slopes of the faint and bright parts of the luminosity dependence, and $e(z)$ is a function describing the cosmological evolution of the characteristic luminosity:
\begin{equation}
\label{eq:ez}
e(z)=\left[\left(\frac{1+z_c}{1+z}\right)^{p_1}+\left(\frac{1+z_c}{1+z}\right)^{p_2}\right]^{-1},
\end{equation}
where $z_c$ is a transition redshift, $L^*/2$ is the bend luminosity at $z=z_c$, and $p_1$ and $p_2$ are the slopes of the redshift dependence. 

A LADE model of this kind is known to be a good approximation of the observed X-ray LF of AGN at redshifts $z\lesssim 3.5$ over the broad range of luminosities, $10^{42}<L_{2-10}<10^{46}$\,erg\,s$^{-1}$ (e.g. \citealt{aird2015}), where $L_{2-10}$ is the luminosity in the 2--10\,keV rest-frame energy band. Moreover, similar models have also been successfully used to describe optical/UV LFs of AGN, namely at $2500\,$\AA\ and in the $b_j$ band, \citep{boyle20002df,palanque2013luminosity,ross2013lf}. 
Therefore, the above model is suitable for our purposes, since our \srg/eROSITA--SDSS sample does not contain high-redshift ($z>2.5$) quasars and low-luminosity ($L_{2-10}<10^{43}$\,erg\,s$^{-1})$ AGN. In what follows, we regard all parameters of the LADE model (we do not consider the normalization and its evolution, as mentioned above) as free ones, albeit with priors, expect for the high-redshift slope $p_2$ in the evolution function, which is difficult to constrain with our data and so we fix it at the best-fit value of $-3.58$ from table 5 in \cite{aird2015}. 
If we take the corresponding prior from table 5 in \cite{aird2015}, we get the same results.

In reality, as has been shown in many X-ray and optically based studies (e.g. \citealt{ueda2014,aird2015}), the cosmological evolution of AGN is more complicated than LADE, but this only becomes important at large redshifts, which are not probed by this study. Therefore, the use of more complicated existing LF models, with their many additional parameters related to cosmological evolution, would unnecessarily complicate our treatment. 

\subsection{Variability}
\label{s:variability}

One of the key elements of our analysis (see Section~\ref{s:formulation}) consists in estimating the contribution of quasar variability to the total scatter of the X-ray--UV luminosity correlation and separating it from the intrinsic scatter. As is well known, AGN are variable both in the X-ray and optical--UV bands on time scales ranging from hours (or even minutes in some Seyfert galaxies) up to many years (essentially up to the Hubble time). We thus need to combine the general knowledge of AGN variability with a proper account of the timing properties of the \srg/eROSITA and SDSS observations on which the current study is based. 

\subsubsection{Timing properties of the data}
\label{s:Timing properties of the data}

Let us first consider how the data used in this study were obtained. As regards the optical band, typically a single photometric measurement was obtained for each quasar during the SDSS imaging, with an exposure time of about 55 seconds per band. Therefore, the UV luminosity estimates ($\luv$) that we use are single-epoch measurements obtained on an observed time scale of $\sim 1$\,minute, which corresponds to a time scale of 
$\sim 30$\,seconds in the quasar rest frame at typical redshifts $z\sim 0.5$--1.5.

As regards the eROSITA data, each quasar was observed during four or five (depending on the celestial position) \srg\ all-sky scans with intervals of six months. Each of such visits lasted between one day (for sources located near the ecliptic plane) and approximately one week (for sources with high ecliptic latitude; note that yet longer exposures are not present in this study because the SDSS footprint does not cover the north ecliptic pole), and in fact was a series of 40-s long exposures separated by four hours (see \citealt{sunyaev2021} for more details about the \srg\ all-sky survey). The X-ray luminosity estimates ($\lx$) that we use are based on source fluxes measured on the map of the Eastern Galactic hemisphere obtained by combining all eROSITA data accumulated during the all-sky survey between Dec. 2019 and Feb. 2022. Therefore, these measurements are affected by quasar variability on several time scales, namely: $\sim 40$\,seconds (single exposure), $\sim 4$\,hours (interval between exposures), $\gtrsim 1$\,day (single visit), $\sim 6$\,months (interval between consecutive scans), and $\sim 2$\,years (duration of the \srg/eROSITA all-sky survey). These times pertain to the observer's frame; in the quasar rest frame, they are shorter by a factor of $(1+z)\sim 2$. 

\begin{figure*}[ht]
    \begin{center}
    	\vspace{\wid}
    	\includegraphics[width=2.2\columnwidth]{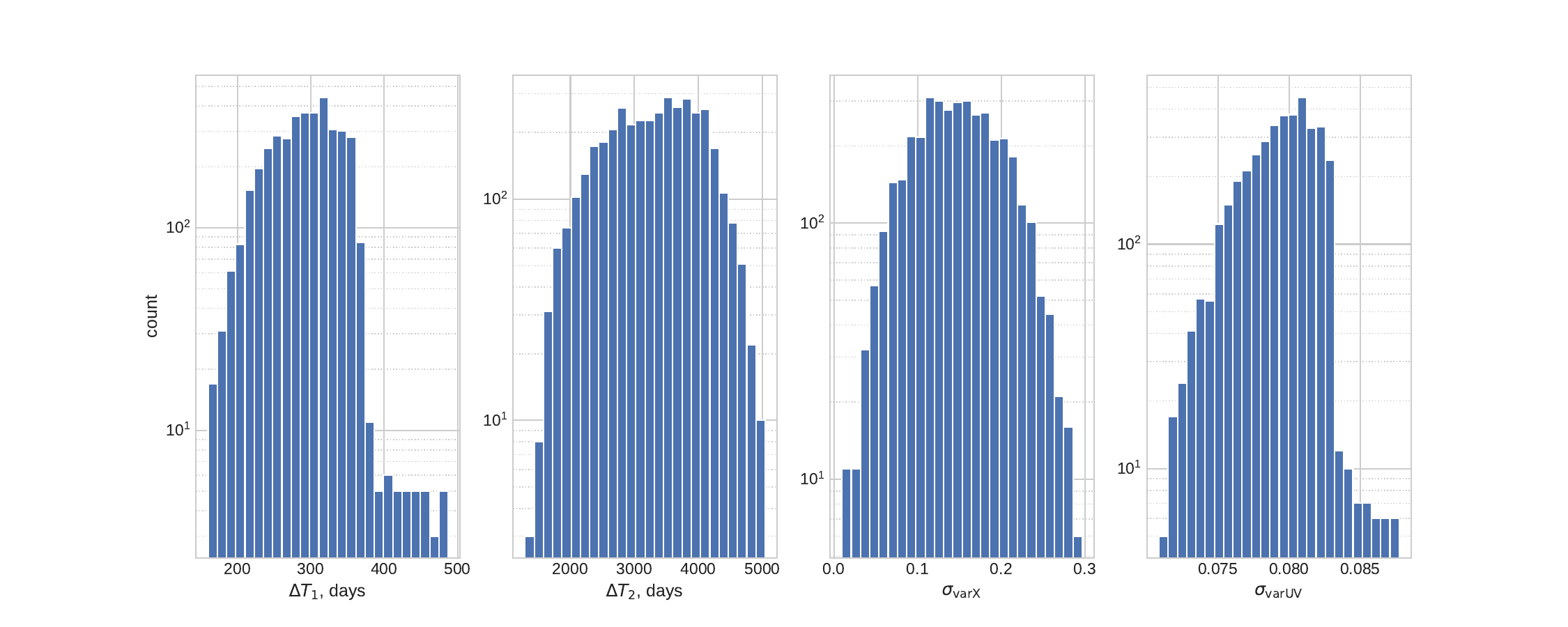}
    	\caption{From left to right: distributions of the rest-frame X-ray (eROSITA) observation periods, rest-frame lags between X-ray and optical photometric (SDSS) measurements and the corresponding distributions of the expected long-term X-ray and short-term UV variability contributions (see main text) to the X-ray--UV luminosity scatter.}
     \label{fig:Xvariability} 
    \end{center}
\end{figure*}

The next important point to consider is that the X-ray data (\srg/eROSITA) were obtained with a significant delay with respect to the optical photometry data (SDSS). Figure~\ref{fig:Xvariability} (second left panel) shows the distribution of lags in the quasar rest frame, $\Delta T_{2}$, between the SDSS and eROSITA observations for our sample. To this end, the X-ray measurement instant for a given quasar is defined as the average between the start time of the first scan and the end time of the last scan (4th or 5th). The rest-frame X-ray--UV lags range between 3.5 and 13.8\,years, and the median $\Delta T_{2}=8.6$\,years. 

\subsubsection{Basic considerations and clarification of the definition of the X-ray--UV luminosity correlation}
\label{s: Basic considerations}

Despite the non-trivial timing properties of the data, a number of well-established observational facts and reasonable assumptions can help us proceed with the treatment of variability. First, we assume that for a given quasar with luminosity $L(t)$ in some energy band at time $t$, the distribution of the logarithms of possible luminosities $l(t+\Delta t)$ after a period of time $\Delta t$ is a Gaussian one centered at $l(t)=\log L(t)$, with variance
\begin{eqnarray}
\sigma^2_{{\rm var}}=\left< \left[l(t+\Delta t)-l(t) \right]^2 \right> \nonumber \\=\left< \left[\log{L(t+\Delta t)-\log{L(t)}}\right]^2\right>\equiv SF^2(\Delta t). 
\label{eq:sf}
\end{eqnarray}
Here, we have introduced a structure function (SF) (or first-order structure function if we follow \cite{simonetti1985} definition), which is often used in variability studies of AGN and other astrophysical objects. In what follows, we will use a `two-argument structure function', $\Sf(\Delta t_1,\Delta t_2)$, which characterizes the variability on rest-frame time scale $\Delta t_2$ of the luminosity averaged over $\Delta t_1 < \Delta t_2$, i.e.
\begin{equation}
\Sf(\Delta t_1,\Delta t_2)\equiv\left< \left[\overline{l(t+\Delta t_2)}-\overline{l(t)} \right]^2 \right>,
\label{eq:gsf}
\end{equation}
where $\overline{l}$ means averaging over $\Delta t_1$.

As known from many studies of Seyfert galaxies and quasars, X-ray and optical variability (e.g. expressed in terms of SF) in a given AGN grows with increasing time scale, and this dependence can to a first approximation be described by a power-law on time scales from days to a couple of decades both in X-rays \citep{vagnetti2011,vagnetti2016,middei2017,prokhorenko2024} and in the optical (see e.g. \citealt{macleod2010,morganson2014}). Therefore, to a first approximation our $\lx$ estimates may be regarded as luminosities averaged over a $\sim 2/(1+z)$\,year period (which corresponds to the full time span of the \srg\ all-sky survey). In contrast, as already mentioned, the adopted $\luv$ values represent much shorter (less than a minute), snapshot estimates of the UV luminosities. 

Recent long-term ($\sim 1$\,year and longer) monitoring (so-called continuum reverberation mapping) campaigns of several AGN have demonstrated that variability in the X-ray and optical--UV bands is usually strongly correlated on time scales of months to years, but much less so on shorter (days to weeks) scales \citep{edelson2019,hernandez2020,cackett2023,lawther2023,edelson2024}. \cite{hagen2024} suggest that long-term X-ray--UV variability correlation can result from accretion-rate fluctuations that propagate inwards through the accretion disc (as in the model of \citealt{lyubarskii1997}) and modulate not only the optical--UV emission produced in the disc but also the X-rays generated in the central hot corona, whereas the complex short-term behaviour is due to additional X-ray variability induced in the corona and due to reprocessing of the X-rays into lower frequency radiation in the disc and, perhaps, in its wind. Although continuum reverberation mapping programs have so far been mostly targeted at nearby, moderate luminosity ($\sim 10^{43}$--$10^{44}$\,erg\,s$^{-1}$ in the X-ray band) Seyfert galaxies, a few similar campaigns have been conducted for higher luminosity AGN. In particular, \cite{arevalo2008} reported on the results of 2.5\,year-long monitoring of quasar MR\,2251$-$178, whose X-ray luminosity ($\sim 5\times 10^{44}$\,erg\,s$^{-1}$, 2--10\,keV) is typical of our \srg/eROSITA--SDSS quasar sample, and found its X-ray and optical light curves to be strongly correlated on times scales of months to years. 

It is thus reasonable to assume that for each quasar there is a `correlation time scale', $\Tc$, shorter than a few months, on which its UV and X-ray emission begin to strongly correlate with each other. More precisely, we assume that intrinsic variability on time scales longer than $\Tc$ `moves' quasars along the primary dependence given by equation~(\ref{eq:lxluv}) on the UV--X-ray luminosity diagram, i.e. if we could average the X-ray and UV luminosities of a given quasar over the same time interval of duration longer that $\Tc$, then the following relation between the mean X-ray and UV luminosities would hold:
\begin{equation}
\langle\lx\rangle=\beta+\gamma\langle\luv\rangle.
\label{eq:avlum}
\end{equation}

This allows us to narrow down the definition of the X-ray--UV luminosity interdependence that we wish to determine. Namely, we seek a relation between UV and X-ray luminosities averaged over a period of time (the same for the X-ray and UV measurements, i.e. they have to be simultaneous) of length $\Delta T_{1}\geq\Tc$. The particular value of $\Delta T_1$ is not important as long as this condition is satisfied. In practice, we define $\Delta T_{1}$ individually for each quasar as the rest-frame time span of its monitoring during the eROSITA all-sky survey, i.e. the time passed between the first and last X-ray observations. This is justified in our case, as $\Delta T_{1}\sim 1$\,year for the \srg/eROSITA--SDSS quasar sample, so that the condition $\Delta T_1>\Tc$ is expected to hold. The distribution of $\Delta T_{1}$ for the \srg/eROSITA--SDSS quasar sample is shown in the leftmost panel on Fig.~\ref{fig:Xvariability}. Accordingly, $\s{intUV}$ and $\s{intX}$ denote the intrinsic scatter of UV and X-ray luminosities with respect to the so-defined primary dependence.

Given this definition, the parameters of the primary dependence and its scatter could be measured directly if the UV and X-ray observations were conducted simultaneously, on the same time scale ($\Delta T_{1}$). Unfortunately, neither condition is met in our case. Therefore, we need to take X-ray and UV variability into account, as described below.

\begin{figure*}[!t]
    	\includegraphics[width=0.9\textwidth]{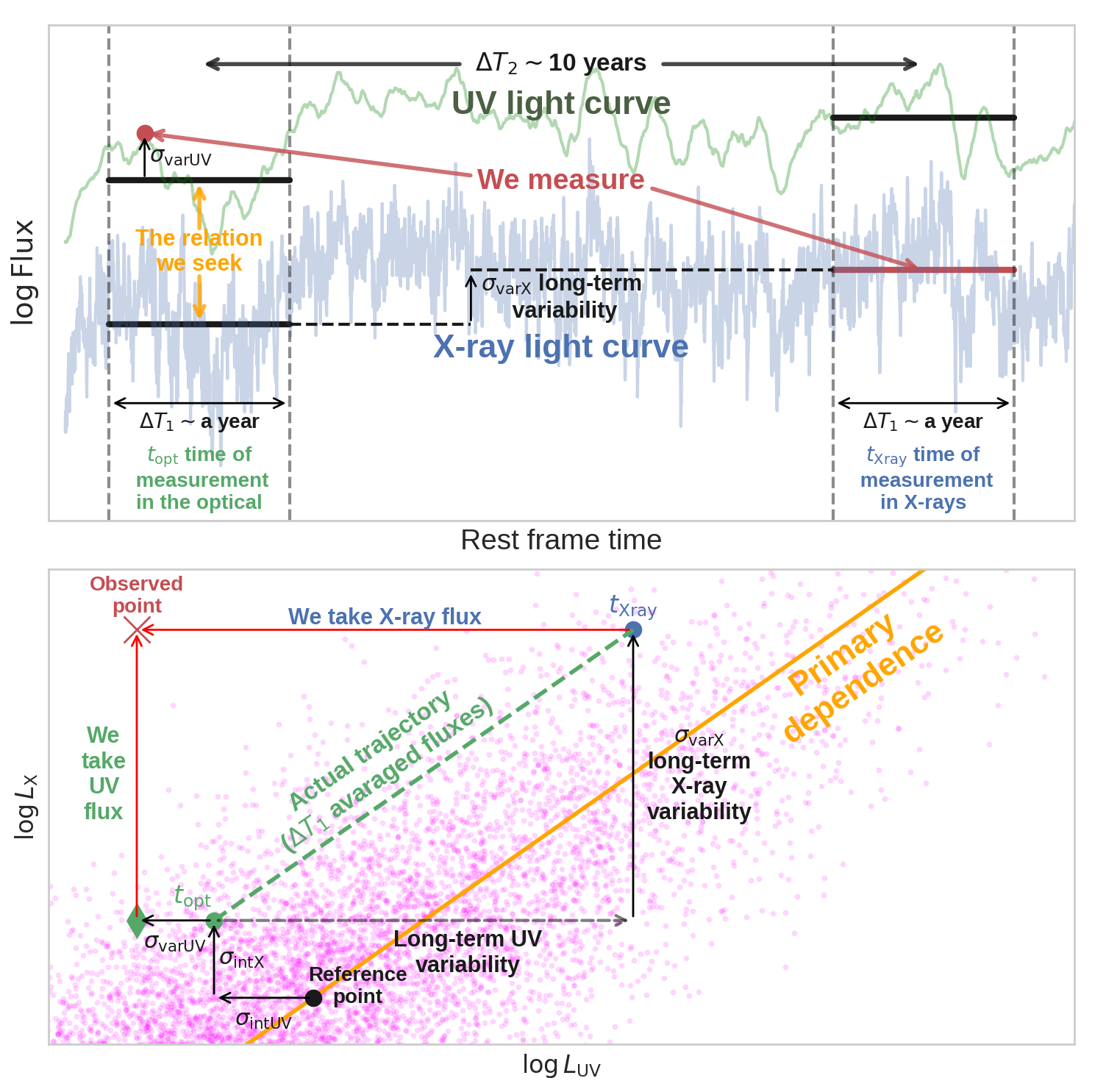}
    	\caption{Illustration of the key aspects of this study. The top panel depicts X-ray (blue) and optical-UV (green) simulated light curves of a quasar and demonstrates what information we can extract from such data. 
        The bottom panel demonstrates the impact of variability and intrinsic scatter on the location of quasars on the X-ray vs. UV luminosity plane. The orange solid line indicates the primary dependence, whose parameters we wish to evaluate. Intrinsic scatter within the quasar population both in X-ray and UV luminosity shifts individual objects away from this dependence, as shown by the black arrows, while variability imposes additional shifts. If X-ray and UV measurements could be done simultaneously and were averaged over sufficiently long intervals (of length $\Delta T_1$), a quasar would travel with time, as a result of variability, along the trajectory shown with the green dashed line. In reality, the X-ray and UV measurements are taken at different instants and the UV (snapshot) measurement is additionally affected by short-term (within $\Delta T_1$) variability. The numerous translucent magenta dots show the intrinsic distribution of the quasar population on this plane at some certain moment in time (taking into account the luminosity function).
        }
     \label{fig:scheme} 
\end{figure*}

We illustrate our approach and the key concepts and assumptions of this study in Fig.~\ref{fig:scheme}. Note that it is made for illustrative purposes only and may be inaccurate in subtleties. The upper panel shows how the X-ray and optical light curves of a typical quasar might look like, as well as the time intervals in which we measure the X-ray and optical fluxes and then infer the corresponding X-ray and UV luminosities using the \srg/eROSITA and SDSS data. We generated the X-ray light curve using the algorithm of \citet{timmer1995}, assuming a broken power-law power density spectrum with slopes of $-1$ and $-2$ and a break time scale of 50\,years. It has a length of 7\,years and a sampling interval of half a day. The UV light curve was obtained by smoothing the X-ray one using a moving mean with uniform weights and a window size of 1\,month (all the quoted time scales being in the quasar's rest frame).

The lower panel of Fig.~\ref{fig:scheme} demonstrates the impact of variability and intrinsic scatter on the location of quasars on the X-ray vs. UV luminosity plane. We adopt the moment of optical flux measurement (which then translates into an estimate of the UV luminosity), $t_{\rm opt}$, as the instant at which we investigate the primary dependence. Note that we could just as well have chosen the time of measurement of the X-ray flux, $t_{\rm Xray}$, for the same purpose.  
We then assume that different quasars have somewhat different intrinsic properties, so that their X-ray luminosities ($\lx$ and $\luv$) averaged over the same $\Delta T_1$ interval around $t_{\rm opt}$ are scattered around the primary dependence, which is illustrated by the green point lying away from the corresponding black point on the primary dependence. However, the X-ray and UV luminosities that we actually infer from observation (here, we ignore the measurement uncertainties for clarity) deviate even more. First, the UV luminosity (the green diamond on the plot) is shifted due to the short-term variability, since the optical flux measured in snapshot SDSS observations fluctuates with variance $\Sigvaruv$ with respect to the $\Delta T_1$-averaged value at $t_{\rm opt}$. Second, the X-ray luminosity, which we estimate from the eROSITA data (the blue point on the plot), shifts due to the long-term variability, as the X-ray flux changes with variance $\Sigvarx$ over the $\Delta T_2>\Delta T_1$ time passed between the optical and X-ray observations. As a result, the observed ($\lx$, $\luv$) pair (the red cross on the plot) lies significantly away from the primary dependence. We have also shown in the bottom panel of Fig.~\ref{fig:scheme} a trajectory that a quasar could follow due to its long-term variability, which we here assumed (for simplicity) to be parallel to the primary dependence.

\subsubsection{Long-term X-ray variability}
\label{s:longterm}

Let us, for convenience, define the $\lx$--$\luv$ correlation at the moment of optical measurement.
We then first need to estimate how the X-ray luminosity of a given quasar changed over the period of time between its observations by SDSS and \srg/eROSITA. Then in equation~(\ref{eq:sigmas}), 
\begin{equation}
\Sigvarx=\Sfx(\Delta T_{1},\Delta T_{2}),
\label{eq:sigvarx}
\end{equation}
(using the definition of the $\Sf$ in eq.~\ref{eq:gsf}).
Note that $\Sfx$ defined here for the monochromatic luminosity at the rest-frame energy of 2\,keV ($\Lx$) is the same as $\Sfx$ for the measured flux in the 0.3--2.3\,keV band, because these quantities are related by a fixed spectral model (power-law with $\Gamma=2$), so that the difference of the corresponding logarithms is the same. 

We have recently conducted a systematic analysis of X-ray variability of quasars \citep{prokhorenko2024} using a similar, but smaller sample of SDSS quasars detected by \srg/eROSITA (and also by \xmm\ for a small part of the sample). The results were presented in terms of $\Sfx(\sim 1\,{\rm day},\Delta T_2)$, with $\Delta T_2$ ranging up to $\sim 10$\,years in the rest frame. We can apply these results to the current analysis using the following relation:
\begin{equation}
\label{eq:SFVar}
\Sfx(\Delta T_{1},\Delta T_{2})=\Sfx(\sim 1\,{\rm day},\Delta T_{2})-2\,\Var_{{\rm X}}(\sim 1\,{\rm day},\Delta T_{1}),
\end{equation}
which is derived in \ref{appendix:C}. Here, $\Var(\Delta t_1, \Delta t_2)$ denotes the variance of luminosity evaluated in $\Delta t_1$ intervals with respect to the average luminosity over a longer period $\Delta t_2> \Delta t_1$.

\cite{prokhorenko2024} demonstrated that the amplitude of X-ray variability on a given time scale is anti-correlated with the black hole mass and accretion rate (Eddington ratio), and with luminosity. Similar trends have been found for optical variability (see e.g. \citealt{arevalo2023}). We adopt from \cite{prokhorenko2024} the following power-law approximation for the dependence of the SF on time lag: $\Sfx (\sim 1\,{\rm day}, \Delta T_{2})=A^2(\Delta T_{2}/{\rm year})^{\beta}$, with the normalization and slope depending on luminosity: $A=-0.111 \log{\frac{L_{2-10}}{\lumU}}+5.145$, $\beta=-0.127\log{\frac{L_{2-10}}{\lumU}}+5.828$, where $L_{2-10}$ is the luminosity in the 2--10\,keV rest-frame energy band. Following \cite{prokhorenko2024}, we estimate $L_{2-10}$ for our quasars from their unabsorbed fluxes in the observed 0.3--2.3\,keV band assuming a power-law spectrum with $\Gamma=1.8$ (which is a typical spectral slope for AGN in the 2--10\,keV energy band).

Therefore, we use for each quasar its specific values of $\Delta T_{1,i}$ and $\Delta T_{2,i}$ and apply the above SF model to calculate $\Sfxi(\sim 1\,{\rm day},\Delta T_{2,i})$ and $\Var_{{\rm X,i}}(\sim 1\,{\rm day},\Delta T_{1,i})$ in equation~(\ref{eq:SFVar}) (see \ref{appendix:C} and, in particular, equation \ref{eq:VSFtPLlim}). Note that we have extrapolated the dependencies from \cite{prokhorenko2024} to calculate $\Var_{{\rm X,i}}(\sim 1\,{\rm day},\Delta T_{1,i})$, i.e. assumed the same slope of $\Sfxi$ on time scales  from $\sim 1$\,day to $\Delta T_{1,i}$ as from $\Delta T_{1,i}$ to $\Delta T_{2,i}$. This is a reasonable assumption, since works based on wider ranges of time scales (see \citealt{vagnetti2016,middei2017}) suggest that this slope does not change significantly. The second right panel of Fig.~\ref{fig:Xvariability} shows the resulting distribution of $\Sigvarx$ for our sample. The median value of $\Sigvarx$ is $0.020$.

\subsubsection{Short-term UV variability}
\label{s:shortterm}

We should now recall that our UV luminosity estimates ($\luv$) are based on snapshot SDSS flux measurements, which are affected by short-term UV variability that is uncorrelated with that in X-rays. Therefore, in accordance with our definition of the X-ray--UV luminosity correlation, we must take into account this short-term UV variability.
Namely, we need to know the expected scatter of $\sim 0.5$-minute UV luminosity measurements within the $\Delta T_{1}$ ($\sim 1$\,year) interval, that is, in equation (\ref{eq:sigmas}),
\begin{equation}
\Sigvaruv=\Var_{{\rm UV}}(\sim 0.5\,{\rm minute},\Delta T_{1}).
\end{equation}

The variance $\Var(\Delta t_1, \Delta t_2)$ can be unambiguously calculated from $\Sf(\Delta t_1, \Delta t_2)$ (see \ref{appendix:C} for details). 

To estimate the UV variability, we use the best-fit parameters of the multidimensional power-law model from the work by \cite{li2018}, who explored the relationship between the SF and $z$, bolometric luminosity, rest-frame wavelength and black hole mass based on a sample of 119,305 SDSS and Dark Energy Camera Legacy Survey (DECaLS) quasars. For simplicity, we ignore all dependencies except the one on wavelength (we adopted a value of $2500$\,\AA). Specifically (see eq. (8) in \citealt{li2018}), we assume that $\log{SF^2_{{\rm UV}}(\tau)}=\log{0.076}-0.551\log(2500\,$\AA$/10000\,$\AA$)+\bigl[0.487+0.172\log(2500\,$\AA$/10000\,$\AA$)\bigl]\log{\tau}-\log{2.5^2}$, where the time lag $\tau$ is measured in years and we have added the last term to convert the SF from \cite{li2018} defined in terms of mag$^2$ to our SF defined in terms of dex$^2$. We ignore the differences between the exposures of SDSS spectroscopy and SDSS and DECaLS photometry. Given the power-law increase of $SF^2$ with time scale and the fact that we consider $SF^2$ on the scale of a year, whereas the characteristic exposure times are days or hours, this will be a minor correction.

The resulting distribution of $\Sigvaruv$ for the \srg/eROSITA--SDSS sample is shown in the rightmost panel of Fig.~\ref{fig:Xvariability}. The median $\Sigvaruv=0.006$. The contribution of short-term UV variability is small compared to that of long-term X-ray variability ($\Sigvarx$). Hence, short-term (on time scales less than $\sim 1$ year) optical--UV variability of quasars plays a minor role in shaping the observed X-ray--UV luminosity relation.

\section{Results}
\label{s:results}

We applied our method to the \srg/eROSITA--SDSS quasar sample. We adopted the mean values and standard deviations of the LADE model of \citet{aird2015} (their Table 5) based on their 'soft band' (0.5--2\,keV) selected AGN sample\footnote{Although \citet{aird2015} did not explicitly exclude type II AGN from their study, as we do, their soft-band sample is expected to be strongly dominated by type I AGN.} as Gaussian priors on the parameters of $\npLF$, except for $g_2$ and $l^*$, for which we adopted much larger standard deviations. This is done because, as was already discussed in Section~\ref{s:luminosityfunction}, the primary LF of our interest is expected to have somewhat different high-luminosity slope and bend luminosity compared to the observed LF. We used broad uniform priors for the parameters of the X-ray--UV luminosity correlation (uniform in logarithm scale for the intrinsic scatter parameters). All the priors are listed in Table~\ref{tab:Priors}. To facilitate comparison of our results for the LF with those of previous studies, in particular with \cite{aird2015}, we define the bend luminosity (see eq.~\ref{eq:LumF}) in the rest-frame 2--10\,keV energy band, namely use the definition
\begin{equation}
l^*=\log \frac{L^*}{{\rm erg\,s}^{-1}{\rm Hz}^{-1}}+17.90,
\label{eq:lstar}
\end{equation}
where $L^*$ is the corresponding monochromatic luminosity (per frequency) at 2\,keV and the constant term corresponds to a power-law spectrum with $\Gamma=2$.

\label{LF}

\begin{table}
	\vspace{2mm}
	\centering
	\vspace{2mm}
	\begin{tabular}{ccc} \hline
	   Parameter & Prior type & Prior specification \\
    \hline
    $\gamma$ & Uniform & from 0 to 3 \\
$\beta$ & Uniform & from $-60$ to 30  \\
$\log\sigma^2_{{\rm intUV}}$ & Uniform & from $-\infty$ to $+\infty$ \\
$\log\sigma^2_{{\rm intX}}$ & Uniform & from $-\infty$ to $+\infty$ \\
$g_1$ & Normal & mean 0.44, std 0.02  \\
$g_2$ & Normal  & mean 3.00, std 2.00  \\
$l^*$ & Normal & mean 44.93, std 1.00 \\
$p_1$ & Normal & mean 3.39, std 0.08  \\
$z_c$ & Normal & mean 2.31, std 0.07 \\
 \hline

 \hline
	\end{tabular}
   \caption{Priors for the model parameters.}
  \label{tab:Priors} 
\end{table}

\subsection{Baseline model}
\label{s:Baseline model}

\begin{figure*}[t]
    	\vspace{6mm}
    	\includegraphics[width=0.95\textwidth]{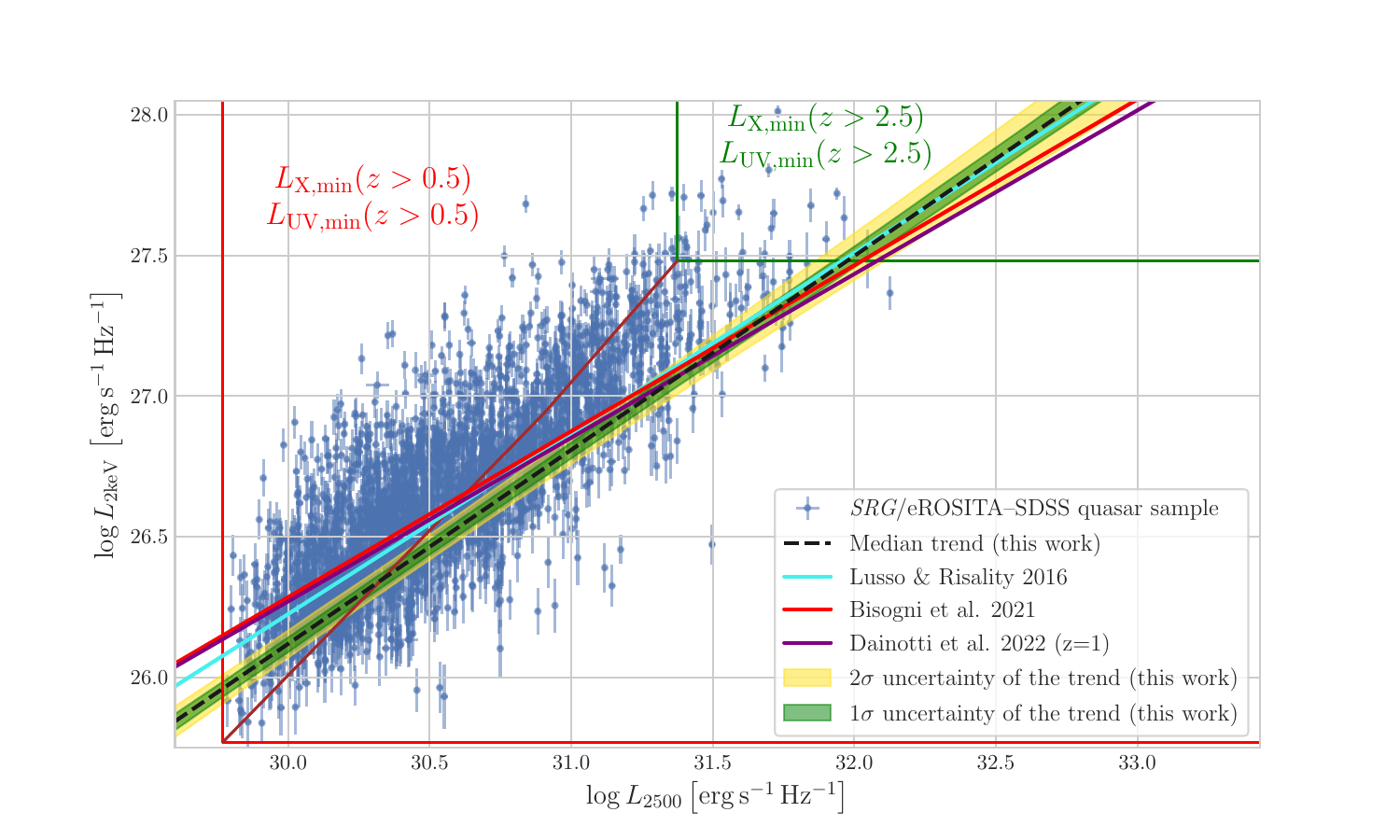}
    	\caption{Scatter plot of the X-ray and UV luminosities with their measurement uncertainties for the \srg/eROSITA--SDSS quasar sample (points with error bars) and the inferred correlation between these quantities for the baseline model: the dashed line corresponds to the median $\gamma$ and $\beta$ values from the posterior distributions and the green and yellow regions show the corresponding 1\,$\sigma$ and 2\,$\sigma$ uncertainty. For comparison, we show (without uncertainties) the $\lx$--$\luv$ relations obtained in previous studies, as indicated in the legend. We also show the regions in the X-ray--UV luminosity space where the current study probes quasars at $z>0.5$ and at $z>2.5$ (the selection region moves along the brown solid line with increasing redshift).
        }
     \label{fig:resPLOT} 

\end{figure*}

\begin{figure*}[t]
    	\includegraphics[width=0.95\textwidth]{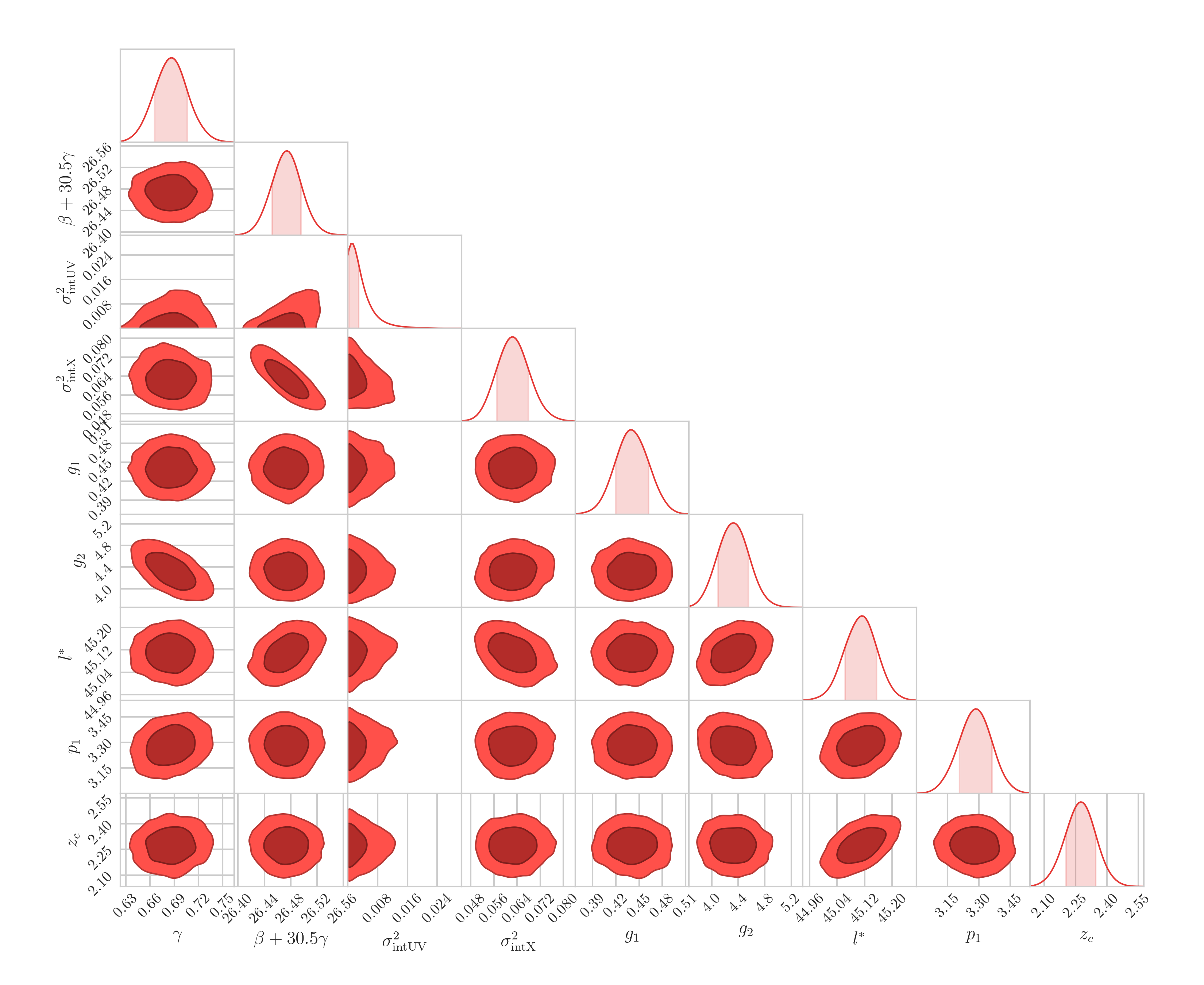}
    	\caption{Posterior distributions of model parameters for the baseline model, including the parameters of the $\lx$--$\luv$ relation and of the LF. The shaded areas of the 1D distributions mark the 68\% smallest credible intervals. The dark and light red areas of the 2D distributions correspond to the 68\% and 95\% smallest credible regions, respectively.}
     \label{fig:resHISTS} 
\end{figure*}

We first consider our `baseline' model, in which there is independent intrinsic scatter in UV and X-ray luminosities ($\s{intUV}$, $\s{intX}$). Figure~\ref{fig:resPLOT} shows the scatter plot of these luminosities for our sample and the inferred correlation between them. The posterior distributions of the model parameters are depicted in Fig.~\ref{fig:resHISTS}, and their key characteristics (mean and median values, standard deviations and percentiles) are presented in Table~\ref{tab:parameters1}. The primary dependence shown in Fig.~\ref{fig:resPLOT} corresponds to the median $\gamma$ and $\beta$ values of the posterior distributions of these parameters, whereas the green and yellow regions around the primary dependence comprise, respectively, [$\luv$, $\lx$] points with $\plx$ lying within the 1$\sigma$ and 2$\sigma$ credible intervals (i.e. 16--84\% and 2.5--97.5\%) of the $\gamma \pluv+\beta$ distribution for a given $\pluv$. 
Namely, for each $\pluv$, we drew a sample of $\left[\gamma,\beta \right]$ pairs from their joint posterior distribution, calculated $\plx=\gamma \pluv+\beta$ and determined the $1\sigma$ and $2\sigma$ credible intervals for $\plx$ based on the resulting $\plx$ distribution.

We tightly constrain the slope of the $\lx$--$\luv$ relation: $\gamma=0.69\pm0.02$ (hereafter, we quote for our results the median values of parameters of interest and the corresponding 1$\sigma$ uncertainties based on the 16--84\% percentile region of the posterior distribution). This value is slightly higher than the slope $\gamma=0.60$--$0.65$ reported by \cite{lusso2016tight} and significantly steeper than the slopes $\gamma=0.59\pm 0.04$ and $\gamma=0.583\pm 0.011$ determined by \cite{bisogni2021chandra} and \cite{dainotti2022}, respectively. As regards the normalization of the $\lx$--$\luv$ relation, we find $\plx(\pluv=30.5)=26.47\pm0.02$, which is somewhat lower than the corresponding values of $26.56\pm0.02$ in \cite{lusso2016tight}, $26.58\pm0.04$ in  \cite{bisogni2021chandra} and $26.57\pm0.03$ in \cite{dainotti2022} (here we have calculated the normalization from \cite{dainotti2022} for the case $z=1$, the characteristic redshift of our sample). This can also be clearly seen from a direct comparison of the X-ray--UV luminosity dependencies in Fig.~\ref{fig:resPLOT}. Hereafter, we utilize the normalization at $\pluv=30.5$, because it is a typical luminosity probed in the current study (the median $\luv$ value in our sample is $30.537$) and in the quoted works. 

\begin{table*}[t]
	\vspace{2mm}
	\centering
	\vspace{2mm}
	\begin{tabular}{ccccccc} \hline
	Parameter & Mean value and & $2.5$\% percentile & 16\% percentile & median & 84\% percentile & $97.5$\% percentile\\ 
 & standard deviation & & & & & \\
 \hline
 \multicolumn{7}{c}{Baseline model}\\
 \hline

$\gamma$ & $0.686,\,0.020$ & $0.65$ & $0.67$ & $0.69$ & $0.71$ & $0.73$ \\
$\beta+30.5\gamma$ & $26.474,\,0.021$ & $26.43$ & $26.45$ & $26.47$ & $26.49$ & $26.52$ \\
$\sigma^2_{{\rm intUV}}$ & $0.003,\,0.003$ & $0.000$ & $0.000$ & $0.002$ & $0.005$ & $0.011$ \\
$\sigma^2_{{\rm intX}}$ & $0.063,\,0.005$ & $0.053$ & $0.058$ & $0.063$ & $0.068$ & $0.074$ \\
$g_1$ & $0.441,\,0.020$ & $0.40$ & $0.42$ & $0.44$ & $0.46$ & $0.48$ \\
$g_2$ & $4.33,\,0.22$ & $3.9$ & $4.1$ & $4.3$ & $4.5$ & $4.8$ \\
$l^*$ & $45.11,\,0.05$ & $45.02$ & $45.06$ & $45.11$ & $45.15$ & $45.20$ \\
$p_1$ & $3.28,\,0.07$ & $3.13$ & $3.21$ & $3.28$ & $3.35$ & $3.42$ \\
$z_c$ & $2.28,\,0.07$ & $2.14$ & $2.21$ & $2.28$ & $2.34$ & $2.41$ \\

 \hline
\multicolumn{7}{c}{Assuming $\s{intUV}=0$}\\
\hline

$\gamma$ & $0.685,\,0.020$ & $0.65$ & $0.67$ & $0.69$ & $0.71$ & $0.73$ \\
$\beta+30.5\gamma$ & $26.471,\,0.018$ & $26.44$ & $26.45$ & $26.47$ & $26.49$ & $26.51$ \\
$\sigma^2_{{\rm intX}}$ & $0.064,\,0.005$ & $0.056$ & $0.060$ & $0.064$ & $0.069$ & $0.075$ \\
$g_1$ & $0.440,\,0.020$ & $0.40$ & $0.42$ & $0.44$ & $0.46$ & $0.48$ \\
$g_2$ & $4.332,\,0.214$ & $3.9$ & $4.1$ & $4.3$ & $4.6$ & $4.8$ \\
$l^*$ & $45.11,\,0.05$ & $45.02$ & $45.06$ & $45.11$ & $45.16$ & $45.20$ \\
$p_1$ & $3.29,\,0.07$ & $3.14$ & $3.22$ & $3.29$ & $3.36$ & $3.44$ \\
$z_c$ & $2.27,\,0.07$ & $2.13$ & $2.20$ & $2.27$ & $2.34$ & $2.41$ \\

 \hline
 \multicolumn{7}{c}{Assuming $\s{intX}=0$}\\
 \hline

$\gamma$ & $0.719,\,0.027$ & $0.65$ & $0.69$ & $0.72$ & $0.75$ & $0.79$ \\
$\beta+30.5\gamma$ & $26.93,\,0.03$ & $26.86$ & $26.90$ & $26.93$ & $26.98$ & $27.02$ \\
$\sigma^2_{{\rm intUV}}$ & $0.119,\,0.013$ & $0.10$ & $0.11$ & $0.12$ & $0.13$ & $0.15$ \\
$g_1$ & $0.439,\,0.021$ & $0.40$ & $0.42$ & $0.44$ & $0.46$ & $0.48$ \\
$g_2$ & $3.272,\,0.12$ & $3.1$ & $3.2$ & $3.3$ & $3.4$ & $3.5$ \\
$l^*$ & $44.0,\,0.6$ & $42.6$ & $43.4$ & $44.1$ & $44.7$ & $45.0$ \\
$p_1$ & $3.38,\,0.08$ & $3.16$ & $3.23$ & $3.32$ & $3.42$ & $3.50$ \\
$z_c$ & $2.31,\,0.07$ & $2.15$ & $2.22$ & $2.29$ & $2.35$ & $2.42$ \\

 \hline
	\end{tabular}
   \caption{Characteristics of the posterior distributions of the model parameters.}
  \label{tab:parameters1} 
\end{table*}

As a byproduct of our analysis, we have also determined the parameters of the quasar primary luminosity function $\pLF$.
As can be seen from Fig.~\ref{fig:resHISTS} and Table~\ref{tab:parameters1}, the derived $\pLF$ parameters are in reasonable agreement with previous determinations of the AGN X-ray LF \citep{ueda2003,hasinger2005luminosity,aird2015,ueda2014}, which adds credibility to our results for the $\lx$--$\luv$ correlation (note that all these studies used the same cosmological parameters as ours). The only exception is the slope of the LF bright end, which we find to be significantly steeper ($g_2=4.4\pm 0.2$) than in previous works: $2.57\pm0.16$ \citep{hasinger2005luminosity}, $2.71\pm0.09$ \citep{ueda2014} and $2.55\pm0.05$ \citep{aird2015}. (Here, we quote the values corresponding to the luminosity-dependent density evolution models in these studies, since the corresponding samples reached into high redshifts, where the LADE model becomes inadequate.) However, this difference is actually expected, since, as discussed above, the observed LF reported by these authors results from  convolution of the $\pLF$ determined here with the intrinsic and variability related scatter of the $\lx$--$\luv$ correlation, which increases the number of quasars with high observed luminosities. We prove this explicitly in \ref{appendix:B}.

Importantly, we find that $\Sigintx$ provides a substantial contribution [more than (long-term) X-ray variability and much more than X-ray flux measurement uncertainties] to the total scatter. In contrast, $\Sigintuv$ just marginally exceeds zero and its contribution to the total scatter is small [as are the contributions of (short-term) UV variability and UV flux measurement uncertainties]. Note, however, that these intrinsic scatter parameters significantly anti-correlate with each other and correlate or anti-correlate with the slope and normalizing constant of the $\lx$--$\luv$ relation. 

\subsection{Intrinsic scatter models}
\label{s:Intrinsic scatter models}

\begin{figure*}[t]
    \begin{center}
    	\vspace{\wid}
    	\includegraphics[width=0.95\textwidth]{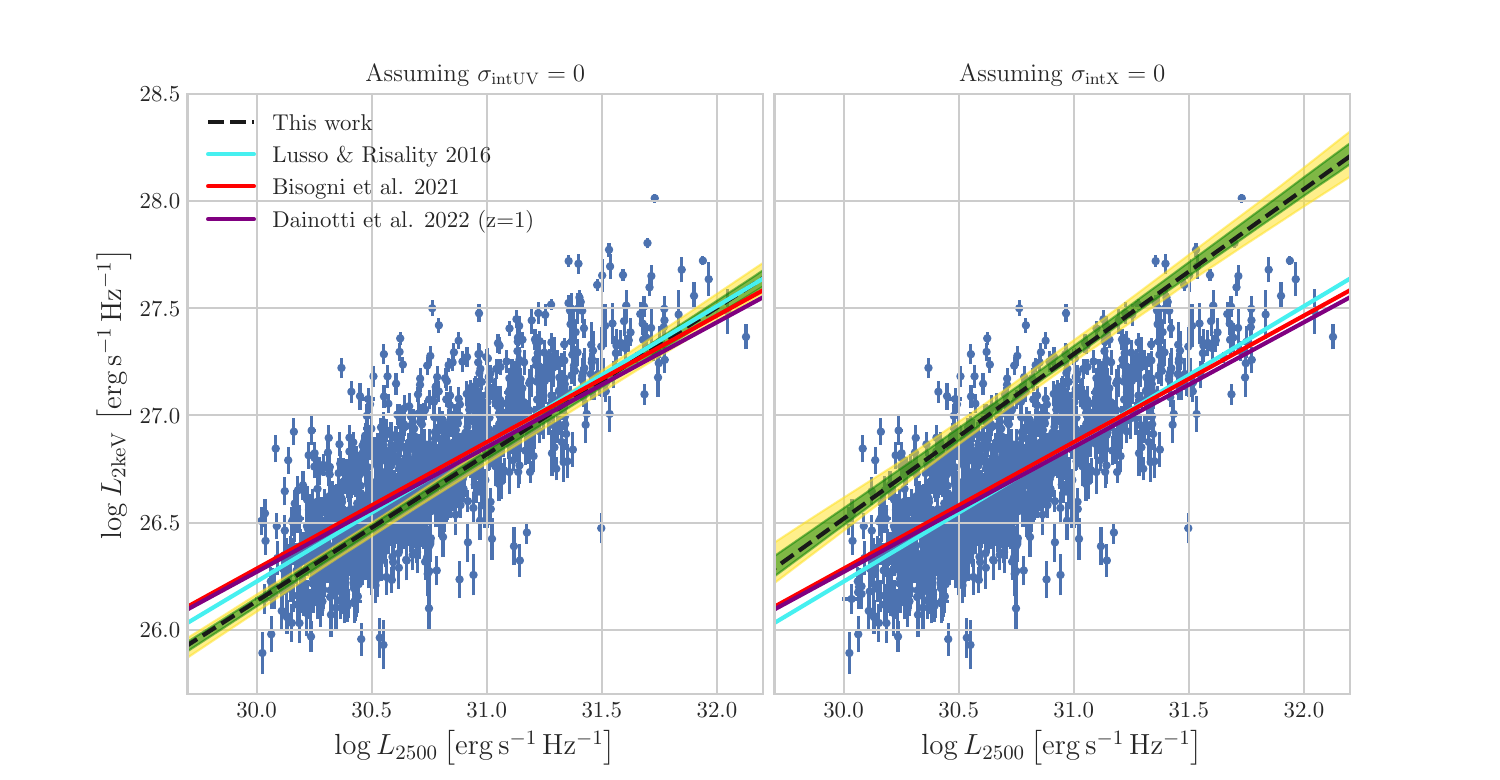}
    	\caption{Same as Fig.~\ref{fig:resPLOT}, but for the $\s{intUV}=0$ and $\s{intX}=0$ models.}
     \label{fig:resPLOTuvx} 
    \end{center}
\end{figure*}

\begin{table}
	\vspace{2mm}
	\centering
	\vspace{2mm}
	\begin{tabular}{cccc} \hline
	   Model & $k$ & $-2\,\ln{\mathcal{L}_{\max}}$ & $AIC$ \\
    \hline
    Both $\s{intX}$ and $\s{intUV}$ are free & 9 & -4157.56 & -4139.56 \\
    Assuming $\s{intUV}=0$ & 8 & -4156.34 & -4140.34 \\
    Assuming $\s{intX}=0$ & 8 & -4091.77 & -4075.77 \\
 \hline

 \hline
	\end{tabular}
   \caption{Comparison of the different intrinsic scatter models.}
  \label{tab:AICtable} 
\end{table}

Essentially all previous studies \citep{lusso2016tight,bisogni2021chandra,dainotti2022} ascribed the entire intrinsic scatter to the X-ray luminosity, i.e. assumed that $\s{intUV}=0$. If we make the same assumption and repeat our analysis, we obtain a slope, $\gamma=0.69\pm0.02$, and a normalization, $\beta+30.5\gamma=26.47\pm0.02$, that are nearly the same as for the baseline model. This is expected, since in that case we obtained $\s{intUV}\approx 0$. The results of this test are presented in the left panel of Fig.~\ref{fig:resPLOTuvx} and in the middle section of Table~\ref{tab:parameters1}.

Alternatively, we can ascribe the entire intrinsic scatter to the UV axis, i.e. assume $\s{intX}=0$. In this case (see the right panel of Fig.~\ref{fig:resPLOTuvx} and the bottom section of Table~\ref{tab:parameters1}), we find nearly the same slope, $\gamma=0.72\pm0.03$, but a much higher normalization, $\beta+30.5\gamma=26.93\pm0.03$, compared to the baseline model. Therefore, the trend line in the $\s{intX}=0$ case is nearly parallel to that in the $\s{intUV}=0$ case but shifted upwards. Moreover, the ratio of $\s{intX}(\s{intUV}=0)$ to $\s{intUV}(\s{intX}=0)$ is equal to $\gamma$ within the uncertainties. 

This behaviour may look surprising at first glance but finds a natural explanation. As demonstrated in \ref{appendix:D}, if the quasars in the Universe obeyed a power-law primary LF, $\npLF(\plx)\propto\pLx^{-\alpha}$, with scatter either only along the X-ray or only along the UV axis, the probability density distribution of quasars on the $\lx$ vs. $\luv$ plane, $f(\lx,\luv|\s{UV},\s{X},\gamma,\beta)$, could be equivalently described by two solutions: one with $\s{uv}=0$ and another with $\s{X}=0$, with the trend line of the latter being translated by $\alpha\sigma^2_{\rm X}(\s{intUV}=0)\ln10$ along the X-ray axis from the trend line of the former and with $\s{X}(\s{UV}=0)/\s{UV}(\s{X}=0)=\gamma$. This behaviour, i.e. the existence of two alternative solutions, could be revealed directly, namely by linear regression $\lx(\luv)$ and $\luv(\lx)$, if one had a volume limited sample of quasars; or, for a flux limited sample, using the maximum likelihood or MCMC based approach developed in our work, which properly takes into account the biases associated with the sample's X-ray and UV flux limits. 

In reality, of course, we are dealing with a more complicated LF, namely with a double power-law model for $\pLF$, hence the $\s{intUV}=0$ and $\s{intX}=0$ solutions are not expected to be parallel to each other anymore. However, to a first approximation (also ignoring the influence of an additional scatter induced by flux measurement uncertainties and variability) we may still expect the $\s{intUV}=0$ and $\s{intX}=0$ solutions to be shifted from each other by an amount that is determined by an `effective' slope of the LF. The measured shift $\beta(\s{intX}=0)-\beta(\s{intUV}=0)\approx 0.5$ falls between the values of 0.06 and 0.62 expected in the single power-law LF scenario for our inferred $\pLF$ slopes $g_1=0.44$ and $g_2=4.3$ below and above $l^*$, respectively. We thus see a reasonable agreement.

We can formally compare the results obtained for the baseline model and for the $\s{intUV}=0$ and $\s{intX}=0$ models using the Akaike information criterion (AIC):
\begin{equation}
\label{eq:AIC}
AIC=2k-2\,\ln{\mathcal{L}_{\rm max}}, 
\end{equation}
where $k$ is the number of estimated parameters in a given model and $\mathcal{L}_{\rm max}$ is the maximum likelihood value, as defined in equation~(\ref{eq:likelihood}). Thus, here we use the standard maximum likelihood estimation method instead of the Bayesian approach. The $\s{intX}=0$ model provides a significantly worse description of the data compared to the baseline model (see Table~\ref{tab:AICtable}). The $\s{intUV}=0$ model is as good as the baseline model in terms of the AIC.

Our derived value for intrinsic dispersion $\Sigintx=0.063\pm 0.005$ corresponds to a standard deviation of $\sim 0.25$\,dex. This is somewhat larger than $\sim 0.2$\,dex estimated by some previous authors \citep{lusso2016tight,chiaraluce2018,bisogni2021chandra, dainotti2022}. The main reason for this difference is likely the following: $\Sigintx$ in our model is defined as the {\it true} dispersion of the two-dimensional luminosity function [$\s{X}^2$ from equation~(\ref{eq:fi}) minus the contribution of variability and measurement uncertainties], rather than as the {\it observed} dispersion (dispersion
relative to a simple regression line based on the observed data points minus the contribution of variability and measurement uncertainties) of a given sample as in the previous studies. As a result, our $\Sigintx$ is expected to be larger because part of the intrinsic distribution of objects is cut off by flux selection in our and other quasar samples. Note that the {\it observed} total dispersion along the X-ray axis, $\sigma^2_{\rm obsX}$, minus the mean $\Sigvarx$ and mean $\Sigmx$ for the \srg/eROSITA-SDSS quasar sample is equal to $0.045$ (equivalent to a standard deviation of $\sim 0.2$\,dex), similar to the previous studies based on other quasar samples. 
Another, perhaps less important, reason for the difference may be related to our sample containing a larger fraction of rare high-luminosity quasars for a given redshift (due to the larger sky coverage by \srg/eROSITA) or exclusion of quasars with inferred significant intrinsic absorption from the previous analyses mentioned above (unlike in our study).

\subsection{Comparison with other methods}
\label{s:Comparison with other methods}

\begin{figure*}[t]
    \begin{center}
   	\vspace{\wid}
    	\includegraphics[width=0.95\textwidth]{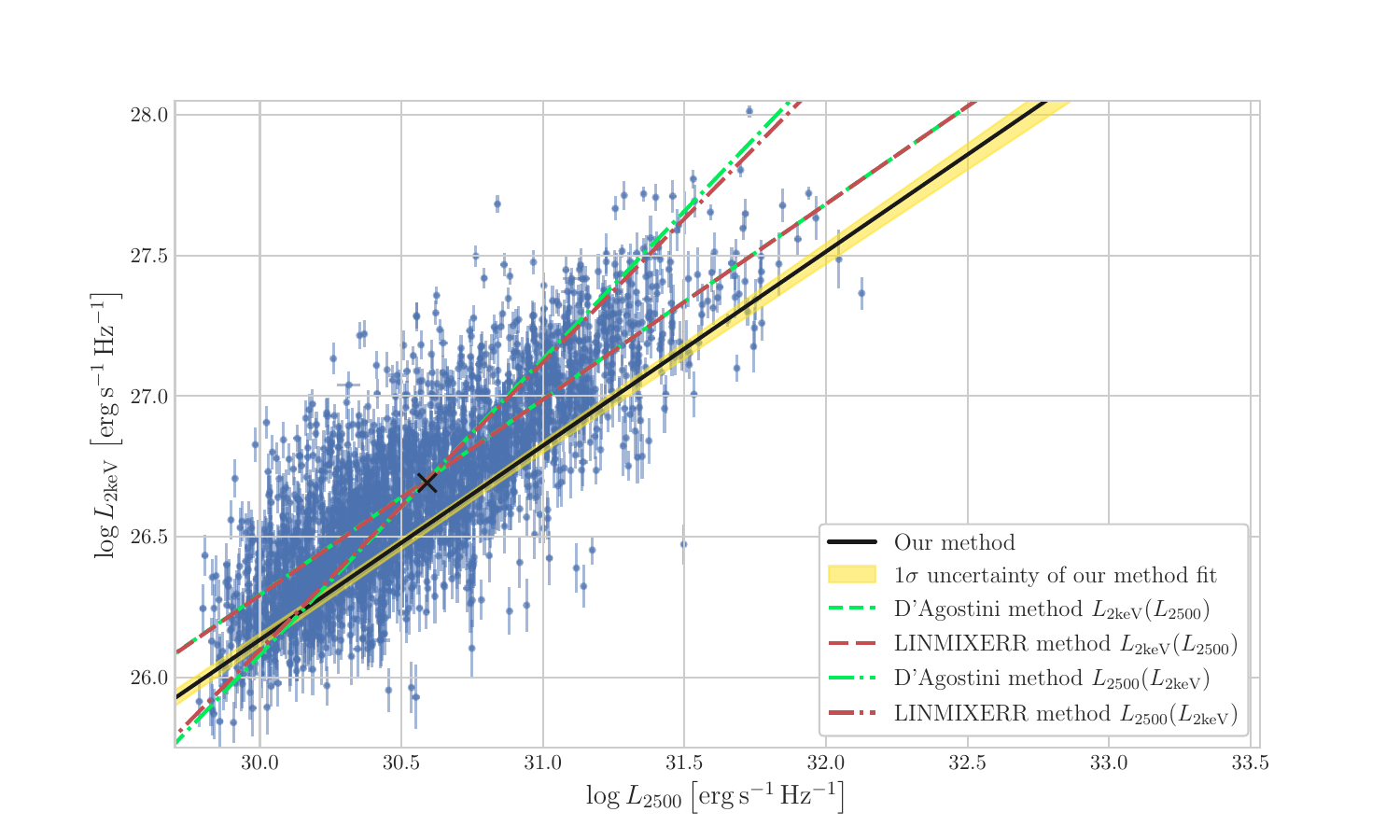}
    	\caption{Comparison of our method with methods from previous works applied to our sample. The solid line corresponds to the median $\gamma$ and $\beta$ values from the posterior distributions and the yellow region shows the corresponding 1$\sigma$ uncertainty for our baseline model. 
        We also show (without uncertainties) the $\lx$--$\luv$ relations for our sample obtained using the D'Agostini and LINMIXERR methods assuming $\Lx(\Luv)$ or $\Luv(\Lx)$ dependencies, as indicated in the legend. The black cross marks the mean values of $\lx$ and $\luv$ for \srg/eROSITA-SDSS quasar sample. }
     \label{fig:resPLOTm} 
    \end{center}
\end{figure*}

As we noted before, in most previous studies (e.g. \citealt{bisogni2021chandra,salvestrini2019,dainotti2022,dainotti2024}), the authors regarded $\lx$ as the ``dependent'' variable, i.e. examined the dependence $\lx(\luv)$. This implies that the problem is one-dimensional (1D), because the likelihood to be minimized is the product of 1D probability densities of $\lx$ for given $\luv$. Thus, it takes the form 
\begin{equation}
\mathcal{L}\propto\prod_i\exp\left(\frac{-[l_{\rm X,i}-\mathcal{F}(l_{\rm UV,i})]^2}{2\delta_{\rm X}^2}\right),
\label{eq:LikeX}
\end{equation}
where $\mathcal{F}$ is an arbitrary (usually assumed to be linear) function and $\delta_{\rm X}^2$ accounts for the scatter along the X-ray luminosity axis (it can include intrinsic scatter, variability and flux measurement uncertainties). One could also consider the opposite case of UV luminosity being the dependent variable, i.e. $\luv(\lx)$, and then
\begin{equation}
\mathcal{L}\propto\prod_i\exp\left(\frac{-[l_{\rm UV,i}-\mathcal{F}(l_{\rm X,i})]^2}{2\delta_{\rm UV}^2}\right),
\label{eq:LikeUV}
\end{equation}
where $\delta_{\rm UV}^2$ defines the scatter along the UV luminosity axis. Hence, the scatter in this approach pertains fully to either X-ray or UV luminosity. \citet{d2005fits} showed that in the 2D case (i.e. considering the probability density of both $\luv$ and $\lx$), assuming constant density along the regression line ($\mathcal{F}$ is linear) one will get the same form as in equations (\ref{eq:LikeX}) and (\ref{eq:LikeUV}) with $\delta^2$ different by a factor of $\gamma$. Since these types of likelihood are equivalent, we refer to them as the D`Agostini method. However, in reality, in the 2D case, the density of quasars is not constant, and the likelihood that accounts for this is irreducible to the simple form of (\ref{eq:LikeX}) and (\ref{eq:LikeUV}).

Other authors, in particular \cite{lusso2016tight,bisogni2021chandra}, used the LINMIXERR method following \cite{kelly2007}, regarding $\lx$ as the ``dependent'' variable (also a 1D case but with a more complex likelihood). This is a linear regression analysis based on the assumptions that (i) the measurement uncertainties are Gaussian with zero mean and known variance, (ii) the intrinsic scatter in the dependent variable about the regression line is Gaussian, and (iii) the intrinsic distribution of the independent variable can be approximated as a mixture of Gaussian functions. The method allows for heteroscedastic and possibly correlated measurement uncertainties.

We have tried to apply these two methods to our \srg/eROSITA--SDSS quasar sample. Specifically, for the D`Agostini method, we used its variant described by \cite{dainotti2022}. Figure~\ref{fig:resPLOTm} compares the results of these analyses, considering $\lx(\luv)$\ and alternatively $\luv(\lx)$, with the results obtained above by our own method (for the baseline model, Sect.~\ref{s:Baseline model}). It turns out that if the X-ray luminosity is assumed to be the dependent variable, then both the D`Agostini and LINMIXERR methods yield nearly the same slope $0.694\pm0.010$ and $0.695\pm0.010$, but significantly higher normalization of the correlation compared to our method. The relations obtained by these methods pass through the center of the cloud of data points, as expected. In contrast, our method is not tied to the observed two-dimensional luminosity distribution (which strongly depends on the flux limits adopted for a given quasar sample). However, if the opposite assumption is made, i.e. the dependence $\luv(\lx)$ is considered, then both the D`Agostini and LINMIXERR methods result in a correlation that is very different from both the result obtained assuming the $\luv(\lx)$ dependence and that obtained by our own method: namely it has a slope of 0.99--1.04. This clearly demonstrates the problem of choosing the `right' dependent variable, which is inherent in the D`Agostini and LINMIXERR methods but is irrelevant for our method.

Why do then our results show such a good agreement with those of \cite{lusso2016tight}, although their analysis was affected by the choice of the dependent luminosity (namely, they adopted $\lx$ to depend on $\luv$)? 
Apparently, this happened because they used a sample with a deep X-ray flux limit, $\sim 10^{-15}$~\flux~in the 0.5--2 \,keV band.
As a result, their sample contained nearly the same number of quasars below and above the true $\lx(\luv)$ dependence, so that the linear regression analysis naturally led to a nearly unbiased result (see also a relevant discussion in \citealt{2012ApJ...757..181S}). 
The fact that we have obtained a similar result using a sample that is characterized by a much higher X-ray flux threshold demonstrates the insensitivity of other method to the choice of sample as regards its X-ray and optical/UV flux limits. 

To further demonstrate this point, we selected from our \srg/eROSITA--SDSS quasar sample a subsample of optically bright objects, namely quasars with $m_g<18.5$ instead of $m_g<19$ (retaining the same X-ray flux limit) and repeated the entire analysis. As demonstrated in \ref{appendix:E}, the slope and normalization of the $\lx(\luv)$ dependence have remained the same within the $1\sigma$ uncertainties (which have of course increased for the smaller sample).

\section{Discussion}
\label{s:discuss}

Below, we briefly discuss several aspects of the results of this study and what could be improved in future work. 

\subsection{Possible origins of the total scatter}
\label{s:Origin of scatter}

Let us recall that we have divided the total dispersion of the $\lx$--$\luv$ correlation into three components: luminosity measurement uncertainties, variability and intrinsic scatter. We relied on the results of previous statistical studies of quasars in estimating the contribution of (long-term) X-ray and (short-term) optical--UV variability, and described the intrinsic dispersion by a pair of free parameters, $\s{intUV}$ and $\s{intX}$. 

Since the current \srg/eROSITA--SDSS sample is composed of fairly bright quasars (in terms of their X-ray and optical--UV fluxes), the X-ray and UV luminosity measurement uncertainties provide minor contributions to the total dispersion. We can formally define the latter with respect to the primary dependence and along the X-ray axis, $\Sigtotx$, as a sum of $\Sigintx$, the mean value of $\Sigvarx$ and the mean value of $\Sigmx$ over the sample. We can similarly define the total dispersion relative to the UV axis, $\Sigtotuv$. We thus find that $\Sigtotx= 0.092\pm0.006$ is composed of $73.4\pm1.8\% $ intrinsic scatter, $22.7\pm1.6\% $ variability (namely, long-term variability) and $3.83\pm0.25\%$ luminosity measurement uncertainties. For $\Sigtotuv=0.009\pm 0.003$, the corresponding fractions are $32^{+16}_{-30}$\%, $67^{+31}_{-16}$\% (namely, short-term variability) and $1.0^{+0.4}_{-0.2}$\%. Therefore, long-term (on time scales of years) variability provides a significant but subdominant contribution to the total dispersion of the $\lx$--$\luv$ correlation for the current sample, allowing us to obtain robust constraints on the intrinsic scatter. 

As was discussed at length in Section~\ref{s:variability}, AGN variability (e.g., characterized by the structure function) grows with increasing time scale. The X-ray flux measurements (\srg/eROSITA) used in our analysis were typically obtained with fairly long rest-frame lags of $\sim 10$\,years with respect to the corresponding optical--UV observations (SDSS). If better synchronized X-ray and optical--UV observations were available for the same quasar sample, the contribution of variability to the total scatter of the $\lx$--$\luv$ correlation would presumably become smaller. It will be interesting to check this prediction using the data of the Dark Energy Spectroscopic Instrument (DESI) survey, which have already provided  \citep{desi2024, DESI_Dawson_2025} new high-quality optical spectra for many \srg/eROSITA quasars, taken nearly at the same time (May 2021--June 2022) as our X-ray observations (Dec. 2019--Feb. 2022).

At the present state of research, we can only speculate about the origin of the intrinsic scatter of the $\lx$--$\luv$ correlation, and in particular why it is aligned with the X-ray luminosity axis. It is natural to expect that at least such key physical parameters as black hole mass and accretion rate (Eddington ratio) as well as, perhaps, spin play important roles in shaping this correlation. It should be possible to address this question quantitatively using the same \srg/eROSITA--SDSS sample, as there are available single-epoch spectroscopy mass estimates for all these quasars. 

Another plausible source of scatter is intrinsic absorption of X-ray and UV emission. We neglected it because our current sample is composed of quasars with broad optical emission lines, however even such objects may be affected by intrinsic absorption, originating near the SMBH and/or in the host galaxy. This, in fact, might be one of the causes of the observed dispersion of the optical--UV $k$-corrections (which reflects the diversity of observed spectral shapes, see the top panel of Fig.~\ref{fig:sampleLmin}). In future work, we plan to address this issue in detail via a careful analysis of the optical--UV spectra of quasars using SDSS and DESI data, which will allow us to clean the sample from strongly reddened spectra and/or correct all spectra for intrinsic extinction. Unfortunately, a similar effort in the X-ray domain is more problematic, as even our current, relatively X-ray bright quasar sample mostly consists of sources from which just a few tens of photons have been detected by eROSITA, which precludes any sensible spectral analysis.

In addition, as has been discussed in the literature (e.g., \citealt{risaliti2011,bisogni2021chandra,signorini2024quasars}), inclination of the accretion disc (usually assumed to be geometrically thin and optically thick, \citealt{shakura1973}) with respect to our line of sight could induce a significant scatter in the observed UV luminosity. Moreover, orientation can similarly affect the observed X-ray luminosity, since the hot corona of the accretion disc is inferred to be moderately optically thick with respect to its X-ray radiation (Thomson optical depth $\tau\sim 1$, e.g. \citealt{malizia2014,lubinski2016,ricci2018}) and there can also be a significant contribution of emission from an optically thick ($\tau\sim 20$) `warm' corona at energies near 2\,keV  (e.g. \citealt{magdziarz1998,kubota2018}). It is difficult to investigate these viewing-angle driven effects as one usually cannot reliably infer the orientation of AGN from observations other than discerning open and obscured directions (type I vs. type II AGN). In order to make progress in this direction, it might be worth using additional information, in particular in the infrared band, which in combination with the optical data could help constrain the opening angle of the obscuring `torus' and the inclination of the accretion disc in a given object.

\subsection{Cosmological aspects}
\label{s:Cosmology}

In this study, we have fixed the cosmological parameters at the commonly adopted values from the literature (see Section~\ref{s:introduction}). However, as these parameters are known with uncertainties, the derived parameters of the $\Lx$--$\Luv$ relation must be affected by them. Postponing a detailed discussion of this crucial issue to our future work, we note that keeping the same \srg/eROSITA--SDSS quasar sample but assuming a different value of the Hubble constant ($H'_0$ instead of $H_0$) will cause all $\lx$ and $\luv$ values to shift by the same constant $2\log{(H_0/H'_0)}$, and, consequently, change the normalization of $\Lx$--$\Luv$ relation and the characteristic scale of the luminosity function in a trivial way: $\beta'=\beta+2\log{(H_0/H'_0)}(1-\gamma)$ and $l^*$$'=l^*+2\log{(H_0/H'_0)}$, whereas the slope ($\gamma$) of the $\Lx$--$\Luv$ relation will not change at all. 

Regarding the potential use of quasars as standardizable candles for cosmology, it is, perhaps, more interesting to study the impact of changing other cosmological parameters, for example $\Omega_M$ within the flat ${\rm\Lambda CDM}$ model. As a first step in this direction, we found out by redoing our entire analysis that if we take $\Omega_M$ values from the interval (0.27, 0.33) (as suggested by current cosmological constraints, \citealt{Kowalski_2008,Planck2018results,Brout_2022,DESICosm,Adame_2025}), all our results remain unchanged within their 1$\sigma$ uncertainties. Similarly, when we considered $\Omega_M$ a free parameter and set a normal prior on it, centered at 0.3 and with a standard deviation of 0.02, the changes proved to be negligible. Note that with these simple tests we are not facing a circularity problem, since we are not trying to estimate cosmological parameters here, but rather fix them or take them from a narrow a priori distribution.

\section{Summary}
\label{s:summary}
We have developed a statistical model of the joint X-ray--UV luminosity distribution of quasars that is based on: (i) intrinsic linear relation (`primary dependence') between the logarithms of these luminosities ($\lx$ and $\luv$), (ii) dispersion along both axes, and (iii) changing density of quasars along the primary dependence (`primary luminosity function'). We 
also estimated the expected contributions of intrinsic variability (resulting from the non-simultaneity of X-ray and optical/UV observations) and flux measurement uncertainties to the total dispersion of the $\lx(\luv)$ dependence, which allowed us to evaluate the residual (`intrinsic') scatter of the latter. Our method properly accounts for all relevant observational biases and for the evolution of quasar space density with redshift.

We started with a sample of quasars from the SDSS DR16Q catalogue cross-matched with the \srg/eROSITA all-sky survey X-ray source catalogue at spectroscopic redshifts between 0.5 and 2.5, with optical and X-ray flux limits of $m_g>19$ and $\Fx>6\times10^{-14}$\flux, respectively. With this bright sample, we achieved nearly 100\% statistical completeness.
After application of several additional filters (removal of blazars etc.), we obtained a `clean' \srg/eROSITA--SDSS quasar sample, containing 2414 objects. We then applied our method to this sample and obtained the following main results:
\begin{itemize} 
\item
The slope of the primary $\lx(\luv)$ dependence $\gamma=0.69\pm0.02$ and its normalization at $\luv=30.5$ is $\lx=26.47\pm0.02$. 
\item
The X-ray intrinsic scatter ($\Sigintx=0.063\pm0.005$) dominates over the UV intrinsic scatter ($\Sigintuv=0.002^{+0.003}_{-0.002}$). We emphasize that this is the {\it true} dispersion of the entire population of quasars rather than the {\it observed} (smaller) dispersion of a given flux limited quasar sample, which was estimated by previous authors. 
\end{itemize}

The derived parameters of the X-ray--UV luminosity relation are in satisfactory agreement with the results of some (but not all) previous studies. However, we emphasize that none of these works accounted for observational selection biases as fully as has been done here, nor did they consider the dispersion of the $\lx$ ($\luv$) dependence for both variables at the same time. As a byproduct of our analysis, we also determined the shape of the X-ray LF of quasars, which is consistent with the results of previous studies. 

In future work, we would like to apply the method developed here to other samples, in particular to DESI optical quasars cross-correlated with \srg/eROSITA X-ray sources. These two surveys were conducted almost simultaneously, which makes it possible to reduce the influence of the variability factor, which impairs the accuracy of our method. We also plan to investigate the influence of SMBH mass and accretion rate, as well as redshift and intrinsic absorption on the X-ray--UV luminosity relation and its dispersion. 

\section*{Acknowledgements}
This work is based on observations with eROSITA telescope onboard
SRG observatory. The SRG observatory was created by Roskosmos in
the interests of the Russian Academy of Sciences represented by its
Space Research Institute (IKI) in the framework of the Russian Federal Space Program, with the participation of Germany. The SRG/eROSITA X-ray telescope was built by a consortium of German Institutes led by MPE, and supported by Deutsches Zentrum f\"{u}r Luft- und Raumfahrt (DLR). The SRG spacecraft was designed, built, launched and is operated by the Lavochkin Association and its subcontractors. The science data are downlinked via the Deep Space Network Antennae in Bear Lakes, Ussurĳsk, and Baykonur, funded by Roskosmos.

This work also made use of data supplied by Sloan Digital Sky Survey IV, provided by the Alfred P. Sloan Foundation.

SP, SS and MG acknowledge support by grant 075-15-2024-647 from the Ministry of Science and Higher Education.

\section*{Data Availability}
The SDSS data used in this article are accessible through the corresponding web pages.
The X-ray data analysed in this article were used by permission of the Russian \srg/eROSITA consortium. The data will become publicly available as part of the corresponding \srg/eROSITA data release along with the appropriate calibration information. 


\medskip

\bibliographystyle{mnras}
\bibliography{corralation}

\appendix

\section{Completeness of the SDSS spectroscopic quasar sample}
\label{appendix:A}

\begin{figure*}[t]
    \begin{center}
    	\vspace{\wid}
    	\includegraphics[width=0.9\textwidth]{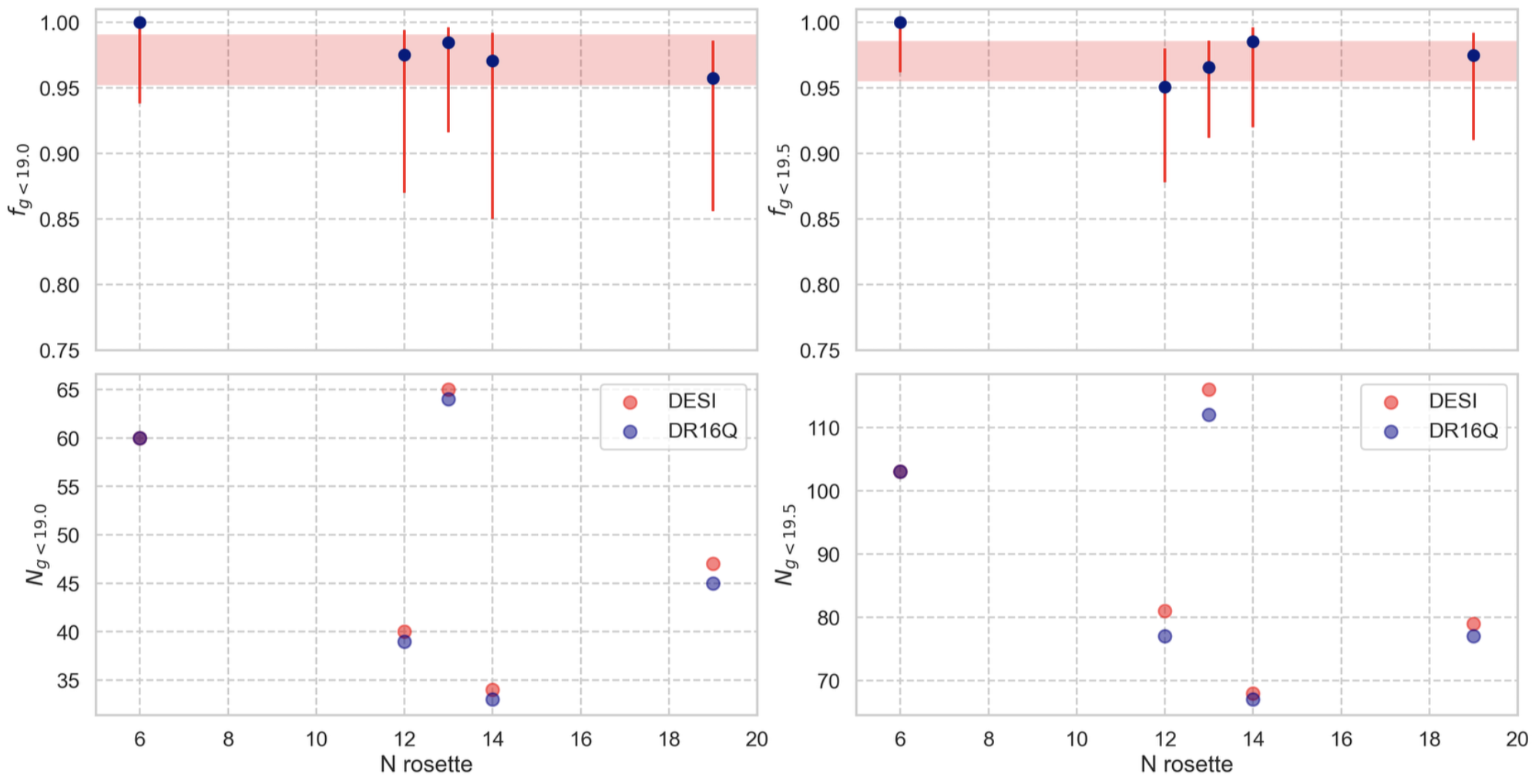}
    	\caption{SDSS DR16Q spectroscopic completeness (top panels) and number of quasars (bottom panels) in different DESI One-Percent survey rosettes within the {\srg}/eROSITA--SDSS footprint. Quasars with $m_g<19$ and $m_g<19.5$ are presented in the left and right panels, respectively. The error bars for the completeness measurements correspond to the 95\% confidence level. The completeness 95\% confidence interval based on all five rosettes is shown by the shaded area for each optical magnitude limit. }
        \label{fig:sp_compl_rosettes} 
    \end{center}
\end{figure*}

Here, we empirically investigate the spectroscopic completeness of the \srg/eROSITA--SDSS sample. Specifically, we are interested in knowing the completeness of SDSS spectroscopy for optical quasars with SDSS dereddened photometric magnitudes $m_g<19$ and $m_g<19.5$ in the redshift interval from 0.5 to 2.5. It is important to note that the SDSS photometric catalogue is highly completed down to much fainter fluxes, namely at $m_g<23.13$ \citep{Abdurrouf_2022}. 

We considered the DESI One-Percent spectroscopic survey \citep{desi2024} as a highly complete reference sample of optical quasars (specifically, we use objects with SPECTYPE=QSO and ZWARN=0 from the DESI EDR SV3 spectroscopic catalogue). The One-Percent survey consists of 20 round areas (`rosettes') of 8.48 square degrees each, where highest spectroscopic completeness has been achieved for targets with $m_g<23$, i.e. much deeper than our adopted brightness limits for the \srg/eROSITA--SDSS sample. Five of these rosettes are fully (more than 98\%) covered by the {\srg}/eROSITA--SDSS footprint, and we use these regions to evaluate the spectroscopic completeness of the SDSS DR16Q catalogue at the photometric limits quoted above. 

\begin{figure*}[t]
    \begin{center}
    	\vspace{\wid}
    	\includegraphics[width=0.9\textwidth]{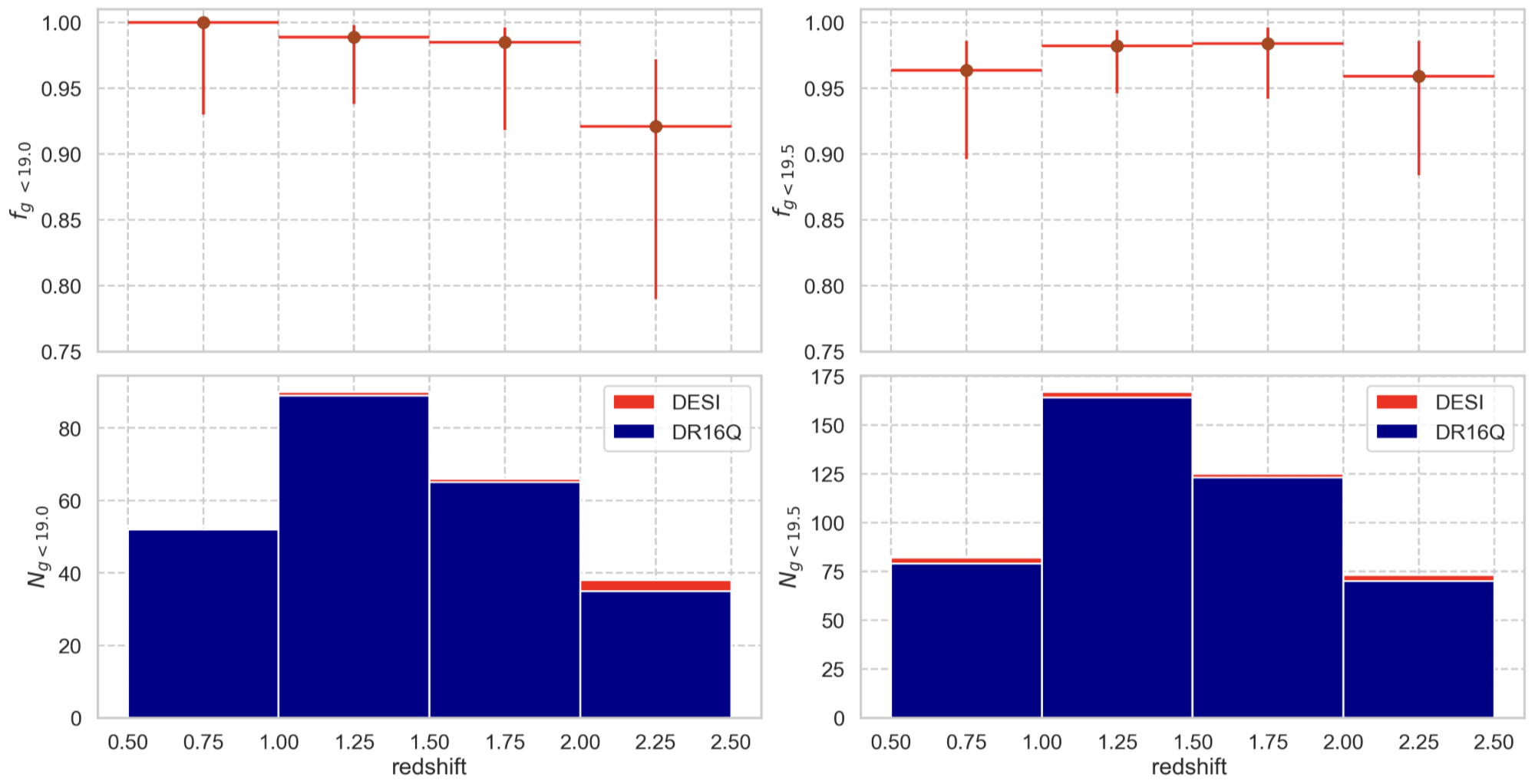}
    	\caption{SDSS DR16Q spectroscopic completeness (top panels) and number of quasars (bottom panels) in different redshift bins. Quasars with $m_g<19$ and $m_g<19.5$ are presented in the left and right panels, respectively. The error bars for the completeness measurements correspond to the 95\% confidence level.}

        \label{fig:sp_compl_z} 
    \end{center}
\end{figure*}

Figure~\ref{fig:sp_compl_rosettes} shows the number (bottom panels) and fraction (top panels) of One-Percent quasars that have also been covered by SDSS spectroscopy, in individual One-Percent rosettes. Quasars with $m_g<19$ and $m_g<19.5$, as defined by SDSS photometry, are presented in the left and right panels, respectively. Also shown is the SDSS spectroscopic completeness based on the sum over the five rosettes. 
In Fig.~\ref{fig:sp_compl_z}, the same statistics is summed over all rosettes and shown as a function of redshift.

The SDSS DR16 catalogue proves to be highly complete (95\% confidence intervals are $\left(94.5,\,97.8\right)$\% and $\left(95.6,\,98.6\right)$\% for the $m_g<19$ and $m_g<19.5$ SDSS photometric cuts, respectively) with respect to the much deeper DESI One-Percent spectroscopic survey across the redshift range of our interest.

\section{Relationship between the primary and observed luminosity functions}
\label{appendix:B}

Let us demonstrate explicitly how the primary LF ($\pLF$, or its normalized variant $\npLF$, which differs from the former by a constant factor, according to eq.~\ref{eq:plf}) corresponds to the observed LF ($NLF$, which differs from $LF$ by the same constant). For $\npLF$, we adopt the double-power law model (eq.~\ref{eq:LumF}) with parameters taken from the range allowed by our baseline $\lx$--$\luv$ correlation model. 

For simplicity, we fix $z=1$, which is a typical redshift for our \srg/eROSITA--SDSS quasar sample. Using equation~(\ref{eq:fdd}), we calculate $NLF$ from the assumed $\npLF$ performing the convolution with $\mathcal{N}(\lx|\plx,\sqrt{\Sigintx+\Sigvarx})$. Here, $\Sigvarx$ depends on $\lx$ as described in Section~\ref{s:longterm}, particularly by equation~(\ref{eq:SFVar}), and we use the median $\Delta T_1$ and $\Delta T_2$ values for our sample. Then we fit the resulting dependence by the double-power law model, with its parameters ($g_1$, $g_2$ and $l_b$) being free, by minimizing deviations between the model and $LF$ (using the least-squares method). Here, $l_b=l^*+\log{e(1)}$, with $e(z)$ and $l^*$ having been defined in equation~(\ref{eq:ez}) and equation~(\ref{eq:lstar}), respectively. To this end, we use 50 X-ray luminosity ($\lx$) bins spaced uniformly over the interval 24.5--28, which is broader than the range effectively probed by our data at $z\lesssim 1$ (see Fig.~\ref{fig:sampleLz}). 

\begin{figure*}[t]
    \begin{center}
    	\vspace{\wid}
    	\includegraphics[width=0.95\textwidth]{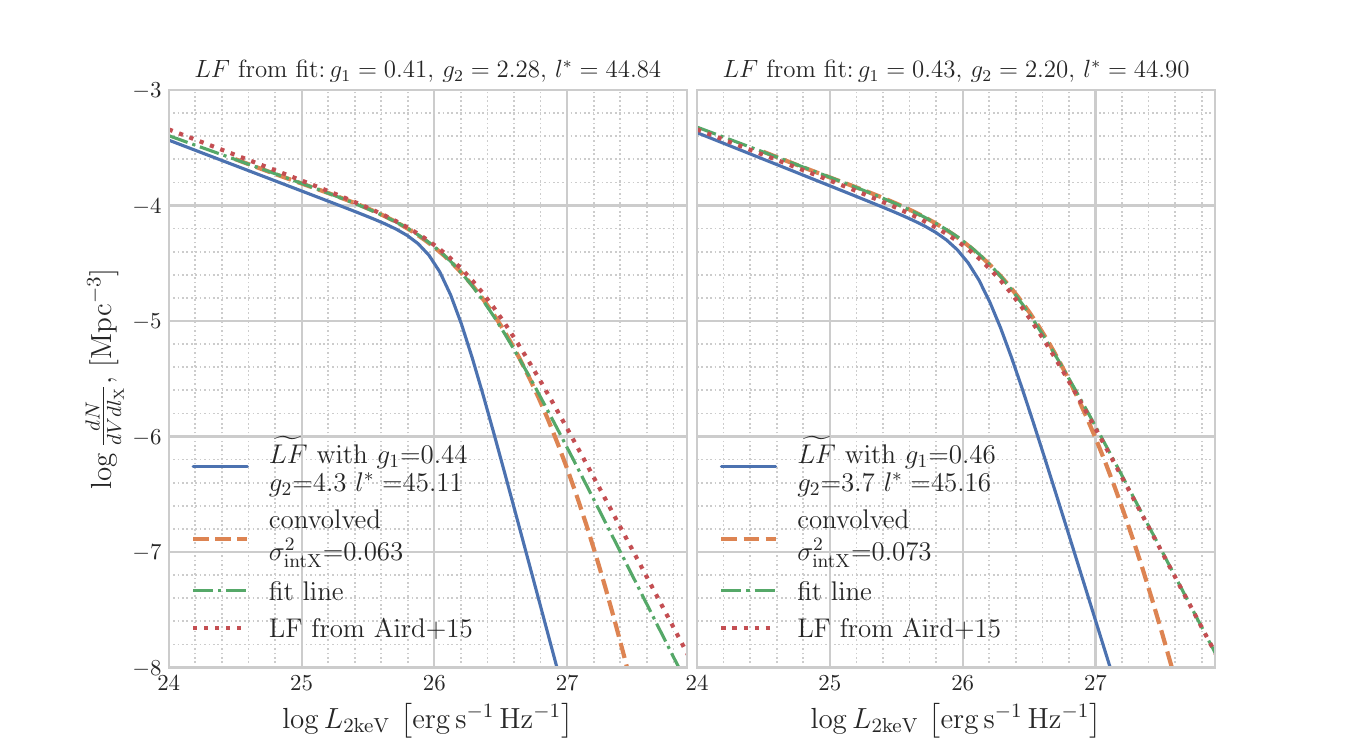}
    	\caption{In each panel, the blue line shows $\pLF$ (see eq.~\ref{eq:TLF}) at $z=1$ with the parameters (of our baseline model) given in the legend; the orange line shows $LF$ at $z=1$ obtained from this $\pLF$ by applying equation~(\ref{eq:fdd}) with intrinsic and variability related scatter (see the main text); and the green line shows the result of fitting $LF$ by the double power-law model (eq.~\ref{eq:LumF}), with the inferred parameters indicated in the header. The left panel shows the results obtained for the median values of $g_1, $$g_2$, $l^*$ and $\Sigintx$ of our baseline model; the right panel for $g_1$ increased by $1\sigma$, $g_2$ decreased by $3\sigma$, $l^*$ increased by $1\sigma$ and $\Sigintx$ increased by $2\sigma$. The red line in each panel shows the soft-band LADE model of \citet{aird2015} (their Table 5) at $z=1$, i.e. with $g_1=0.44\pm0.02$, $g_2=2.18\pm0.04$, $l^*=44.93\pm0.03$, $p_1=3.39\pm0.08$, $p_2=-3.58\pm0.26$ and $z_c=2.31\pm0.07$.}

        \label{fig:LFscatt} 
    \end{center}
\end{figure*}

First, we fixed the parameters of $\pLF$ at the 0.5-quantile (median) values of our baseline model (listed in the top section of Table~\ref{tab:parameters1}). The resulting $\pLF$, $LF$ and the corresponding fit of the latter along with the inferred parameter values are presented in the left panel of Fig.~\ref{fig:LFscatt}. Both $\pLF$ and $LF$ have been normalized to the best a posteriori value (mind that it has uncertainty) of the soft-band LADE model from \citet{aird2015} (see their Table 5) at $z=1$ and $\lx=26.1$ (which corresponds to a luminosity of $10^{44}$\,\lum\ in the 2--10\,keV energy band) and thus have the meaning of quasar space density per $\lx$. We see a fairly good but not perfect correspondence between the results of \citet{aird2015} and ours. Namely, the $g_1$ value inferred from our fit to $LF$ is smaller by $\sim 2\sigma_{\rm A}$, where $\sigma_{\rm A}$ is the uncertainty of this parameter in \citet{aird2015}, while our best-fit $g_2$ and $l^*$ values deviate from theirs each by $\sim 3\sigma_{\rm A}$. However, it is possible to bring our modeled $LF$ to better agreement with the result of \citet{aird2015} by slightly changing the parameters of the adopted $\pLF$. Namely, if we increase $g_1$ by $1\sigma$, decrease $g_2$ by $3\sigma$, increase $l^*$ by $1\sigma$ and increase $\Sigintx$ by $2\sigma$ (where $\sigma$'s are the corresponding uncertainties in our baseline model, see again the top section of Table~\ref{tab:parameters1}), i.e. within the ranges marginally allowed by our baseline model posterior distributions, then we can reproduce the LF of \citet{aird2015} very well, as demonstrated in the right panel in Fig.~\ref{fig:LFscatt}. 

We conclude that the results of the current study are in satisfactory agreement with the work of \cite{aird2015} as regards the AGN LF, given that we have used many assumptions in our analysis that were irrelevant to theirs (in particular, we assumed $\Sigintx$ to be independent of $z$ and $\lx$ and a constant $\Sigvarx$, both of which might not be true). Furthermore, we see that $LF$ can be fairly well described by a double-power law. However, the observed deviations at high luminosities indicate that this kind of model may not be a perfect choice for the primary LF.

\section{Relationship between variance and structure function}
\label{appendix:C}

Consider a weakly non-stationary variability process (see \citealt{vaughan2003}) and its individual realization -- a light curve of length $\Delta t_2$ that consists of $N_1$ consecutive measurements $f_i$ of length $\Delta t_1=\Delta t_2/N_1$ (without gaps between the measurements, i.e. we collect entire fluence in $\Delta t_2$). We can calculate the squared deviation of the measurements from the mean value $\overline{f}$ for this light curve. Let us define the variance on timescales from $\Delta t_1$ to $\Delta t_2$ to be the arithmetic mean of the squared deviation: $\Varj{}= \frac{1}{N_1}\sum_{i=1}^{N_1} \left(f_i- \overline{ f}\right)^2$. 
We are interested in the expected value of $\Varj{}$ for the stochastic process, $\Var(\Delta t_1,\Delta t_2)=\left<\Varj{}\right>_{{\rm lc}}$, i.e. the average over all possible realizations (light curves) of this process. 

Let us also define a two-argument structure function $\SfO$ for the light curve as the mean squared difference between $f_i$ separated by time lag $\tau$, such that $\Delta t_1\leq\tau\leq (\Delta t_2-\Delta t_1)$ and $\tau/\Delta t_1$ is a natural number: $\SfO(\Delta t_1,\tau)= \frac{1}{N_p}\sum_{i}\left(f_i-f_{i+\tau/\Delta t_1}\right)^2$, where the summation is performed over all measurement pairs separated by $\tau$ and $N_p=(\Delta t_2-\tau)/\Delta t_1=N_1-\tau/\Delta t_1$ is the number of such pairs. Accordingly, we define $\Sf=\left<\SfO\right>_{{\rm lc}}$ as the expected value of $\SfO$ over all possible realizations of the stochastic process.

Now, a long light curve of length $\Delta t_3 \gg \Delta t_2$, consisting of consecutive measurements $f_i$ of length $\Delta t_1$, can be divided into consecutive segments of length $\Delta t_2$. Then, the total number of measurement pairs separated by $\tau$, such that $\Delta t_2\leq\tau\leq (\Delta t_3-\Delta t_2)$ and $\tau/\Delta t_2$ is a natural number, in the long light curve is $N_p=(\Delta t_3-\tau)/\Delta t_1=(\Delta t_2/\Delta t_1)(\Delta t_3-\tau)/\Delta t_2=N_1 N_{p2}$, where $N_{p2}\equiv (\Delta t_3-\tau)/\Delta t_2$ is the total number of pairs of these segments. Let us also define $n=\tau/\Delta t_1$, $m=\tau/\Delta t_2$ and for $k$-th light curve of length $\Delta t_2$ we denote its mean and variance as $\overline{f}_{k}$ and $\Varj{k}= \frac{1}{N_1}\sum_{j} \left(f_j- \overline{ f}_k\right)^2$. Then, we have
\begin{align}
&\SfO(\Delta t_1,\tau)=\frac{1}{N_p}\sum_{i=1}^{N_p}\left(f_i-f_{i+n}\right)^2&\nonumber\\
&=\frac{1}{N_p}\sum_{k=1}^{N_{p2}} \sum_{j=(k-1) N_1+1}^{k N_1} \left[ \left(f_{j}-\overline{ f}_k\right)+\left(\overline{f}_{k+m} -f_{j+n}\right)+\left(\overline{f}_{k}-\overline{f}_{k+m}\right)\right]^2&\nonumber\\
&=\frac{1}{{N_{p2}}}\sum_{k=1}^{N_{p2}}\left(\Varj{k}+\Varj{k+m}\right)+\Sfj{}(\Delta t_2,\tau)&\label{C1}\\
&+\frac{2}{N_{p2}}\sum_{k=1}^{N_{p2}} \sum_{j=(k-1) N_1+1}^{k N_1}\frac{\left(f_{j}-\overline{ f}_k\right)\left(\overline{f}_{k+m} -f_{j+n}\right)}{N_1}& \label{C2}\\
&+\frac{2}{N_{p2}}\sum_{k=1}^{N_{p2}} \left[\left(\overline{f}_{k}-\overline{f}_{k+m}\right)\sum_{j=(k-1) N_1+1}^{k N_1}\left(\frac{f_{j}-\overline{ f}_k}{N_1}+\frac{\overline{f}_{k+m} -f_{j+n}}{N_1}\right)\right].& \label{C3}
\end{align} 

The term (\ref{C3}) obviously vanishes due to the definitions of arithmetic mean $\overline{f}_k$ and $\overline{f}_{k+m}$. Let us average $\SfO(\Delta t_1,\tau)$ over many long light curves. Due to the linearity of the expected value, it will be the sum of the average values of (\ref{C1}) and (\ref{C2}). Since we have assumed that the light curve has no `memory' (the $i$-th value of the light curve is not affected by the $(i+n)$-th value), the term (\ref{C2}) vanishes upon averaging, and we finally get:
\begin{align}
\label{eq:SFcon}
\Sf(\Delta t_1,\tau)=2\Var(\Delta t_1,\Delta t_2)+\Sf(\Delta t_2,\tau).
\end{align} 

Let us now establish the relation between $\Sf(\Delta t_1,\tau)$ (defined for $\Delta t_1\leq\tau\leq(\Delta t_2-\Delta t_1)$)
and $\Var(\Delta t_1,\Delta t_2)$. 
We can compose a total of $N=N_1 \left(N_1-1 \right)/2$ measurement pairs from our light curve of length $\Delta t_2$.
On the one hand, we can consider the mean squared difference between two arbitrary measurements from the light curve $f_i$ and $f_k$, and get:
\begin{align}
\label{eq:var2}
&\sum_{k>i}\frac{\left(f_i-f_k\right)^2}{N}=\sum_{k\ne i}\frac{\left(f_i-f_k\right)^2}{2N}=\sum_{i=1}^{N_1}\sum_{k=1}^{N_1}\frac{\left(f_i-\overline{f}+\overline{f}-f_k\right)^2}{N_1\,(N_1-1)}&\nonumber\\
&=\sum_{i=1}^{N_1}\frac{\left(f_i-\overline{f}\right)^2}{N_1-1}+\sum_{k=1}^{N_1}\frac{\left(f_k-\overline{f}\right)^2}{N_1-1}-2\sum_{i=1}^{N_1}\sum_{k=1}^{N_1}\frac{\left(f_i-\overline{f}\right)\left(f_k-\overline{f}\right)}{N_1\,(N_1-1)}&\nonumber\\&=2\,\frac{N_1}{N_1-1}\Varj{}(\Delta t_1,\Delta t_2)-2\sum_{i=1}^{N_1}\frac{f_i-\overline{f}}{N_1-1}\sum_{k=1}^{N_1}\frac{f_k-\overline{f}}{N_1}&\nonumber\\&=2\,\frac{N_1}{N_1-1}\,\Varj{}(\Delta t_1,\Delta t_2),&
\end{align} 
where the definition of $\overline{f}$ was used to cancel the last term in the expression preceding the final one. 
On the other hand, each pair of measurements is characterized by the time interval $\tau_j=j\Delta t_1$, where $j$ can take values from 1 to $N_1-1$. Then, for a given $\tau_j$ there are $p_j=N_1-j$ pairs, and we can write
\begin{align}
\label{eq:var2h}
\frac{\sum_{k>i}\left(f_i-f_k\right)^2}{N}=
\frac{\sum_{j=1}^{N_1-1}\sum_{i=1}^{p_j}\left(f_i-f_{i+j}\right)^2}{\sum_{j=1}^{N_1-1}p_j}
=
\frac{\sum_{j=1}^{N_1-1}p_j \Sfj{j}}{\sum_{j=1}^{N_1-1}p_j}.
\end{align} 
Here, $\Sfj{j}\equiv\frac{\sum_{i=1}^{p_j}\left(f_i-f_{i+j}\right)^2}{p_j}$, and it is related to $\Sf(\Delta t_1,\tau_j)$ as $\Sf(\Delta t_1,\tau_j)=\left<\Sfj{j}\right>_{{\rm lc}}$. Therefore, equations (\ref{eq:var2}) and (\ref{eq:var2h}) give:
\begin{align}
\label{eq:VSF0}
\frac{\Delta t_2}{\Delta t_2-\Delta t_1}\,\Varj{}(\Delta t_1,\Delta t_2)=
\frac{\sum_{j=1}^{\Delta t_2/\Delta t_1-1}\left(\Delta t_2-j\Delta t_1\right) \Sfj{j}}{\left( \Delta t_2-\Delta t_1\right)\Delta t_2/\Delta t_1},
\end{align} 
which, due to the linearity of the expected value over all possible realizations of the process yields
\begin{align}
\label{eq:VSFj}
\Var(\Delta t_1,\Delta t_2)=
\frac{\sum_{j=1}^{\Delta t_2/\Delta t_1-1}\left(\Delta t_2-j\Delta t_1\right) \Sf(\Delta t_1,j\,\Delta t_1)}{\Delta t_2^2/\Delta t_1}.
\end{align} 

This proof reflects the fact that the structure function determines the contribution of measurement pairs with a given time lag to the variance. Note that we would come to the same result if we randomly chose $N'$ times first $f_i$ and then $f_k$ from the set of measurements. That is, the proportion between the total number of measurement pairs $N'$ and the number of measurement pairs with particular $\tau_j$,  for all $j$, does not change when $N' \to \infty$.

In the limit $\Delta t_2 \gg \Delta t_1$,
\begin{align}
\label{eq:VSFt}
\Var(\Delta t_1,\Delta t_2)\approx \frac{1}{2}\int_{0}^{\infty} T(0,0,\Delta t_2,\tau)\Sf(\Delta t_1,\tau) d\tau\nonumber\\
=\int_{0}^{\Delta t_2} \frac{\left(\Delta t_2-\tau\right)}{\Delta t_2^2} \Sf(\Delta t_1,\tau) d\tau,
\end{align} 
where $T(a,c,b,x)$ is the probability density function of the triangular distribution.

For example, in the case of a power-law structure function, $\Sf(\Delta t_1,\tau)=A \tau^\beta$ ($\beta\geq0$), equation (\ref{eq:VSFt}) gives

\begin{align}
\label{eq:VSFtPLlim}
\Var(\Delta t_1,\Delta t_2)\approx\frac{A\Delta t_2^\beta}{\left(\beta+1\right)\left(\beta+2\right)}.
\end{align} 

\begin{figure}[t]
    \begin{center}
    	\vspace{\wid}
    	\includegraphics[width=1.1\columnwidth]{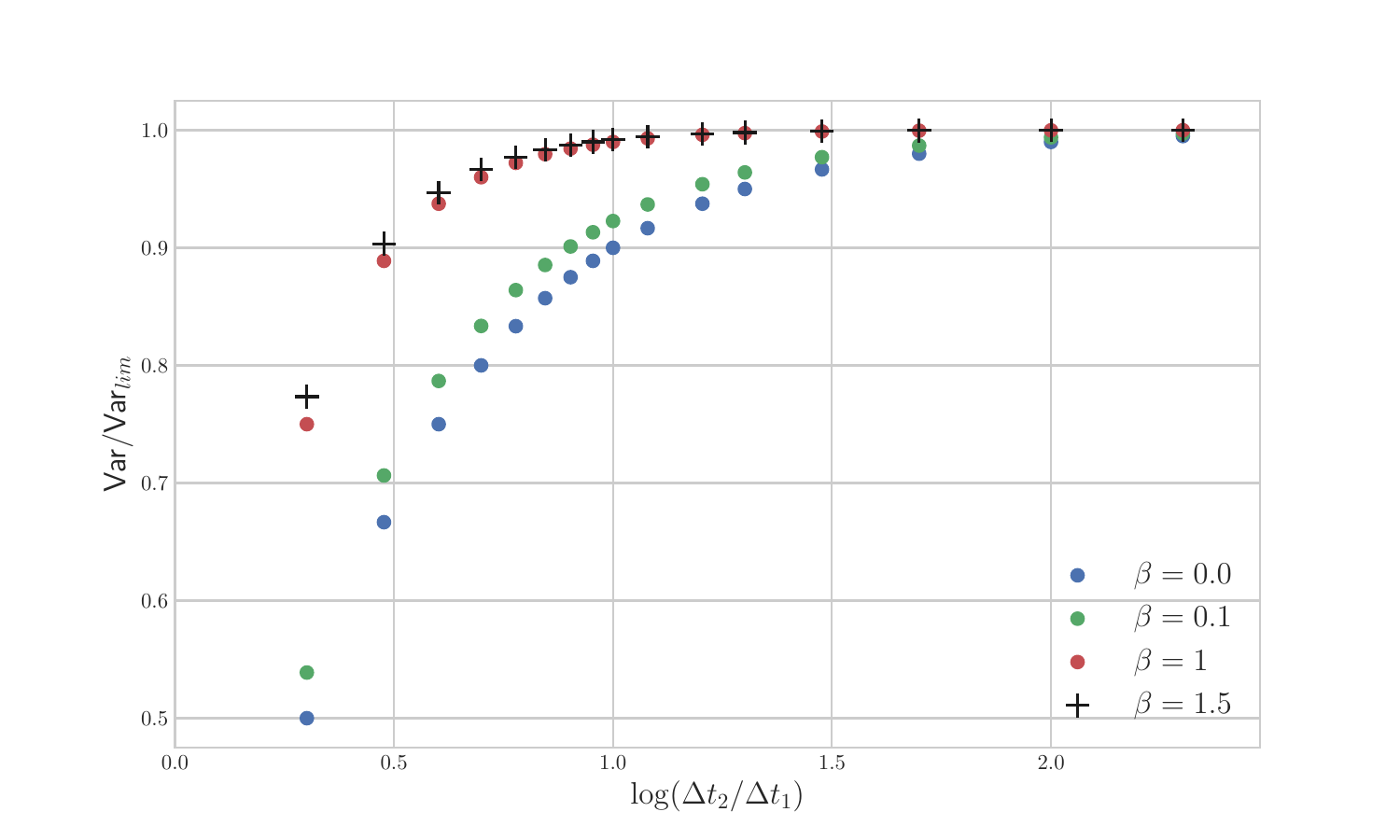}
    	\caption{Ratio of $\Var$  given by equation~(\ref{eq:VSFj}) and its approximation $\Var_{\rm lim}$ given by equation~(\ref{eq:VSFtPLlim}) for a power-law structure function $\Sf(\Delta t_1,\tau)=A\tau^\beta$ as a function of $\Delta t_2/\Delta t_1$, for different slopes (indicated by different colours).}

        \label{fig:SF2vsVar} 
    \end{center}
\end{figure}

Figure~\ref{fig:SF2vsVar} compares $\Var(\Delta t_1,\Delta t_2)$ calculated via equation~(\ref{eq:VSFj}) with its approximation given by equation~(\ref{eq:VSFtPLlim}) for different values of $\Delta t_2/\Delta t_1$ and different slopes $\beta$ of the power-law structure function. For $\Delta t_2/\Delta t_1\geq10$, the approximate expression is accurate within 11\% for all probed slopes.

Thus, knowing the structure function on some time scales, one can readily estimate the variance on these time scales. Moreover, using equation~(\ref{eq:SFcon}) it is also possible to evaluate the structure function on a different time scale. 

\section{The case of a power-law luminosity function and scatter along either the UV or X-ray axis.}
\label{appendix:D}

It is instructive to consider the special case for the probability density $f(\lx, \luv|\s{UV},\s{X},\gamma,\beta)$ (defined in Section~\ref{s:formulation}) where the normalized primary LF has a power-law shape: $\npLF(\plx)\propto \pLx^{-\alpha}\propto 10^{-\alpha\plx}$. If we also assume that there is no scatter in the primary X-ray--UV luminosity relation along the X-ray axis, i.e. $\s{X}=0$, then equation~(\ref{eq:fi}) simplifies to:
\begin{align}
\label{eq:fipluv}
&f(\lx, \luv|\s{UV},\s{X}=0,\gamma,\beta)\propto\mathcal{N}(\luv|(\lx-\beta)/\gamma,\s{UV})\, 10^{-\alpha \lx }&\nonumber\\&\propto\exp\left(-\alpha \lx \ln10-\frac{\left[\gamma\luv-\lx+\beta\right]^2}{2\Siguv\gamma^2}\right).&
\end{align}
If, instead, we assume $\s{UV}=0$, we get:
\begin{align}
\label{eq:fiplx}
&f(\lx, \luv|\s{UV}=0,\s{X},\gamma,\beta)\propto\mathcal{N}(\lx|\gamma\luv+\beta,\s{X})\, 10^{-\alpha (\gamma\luv+\beta)}&\nonumber\\&\propto\exp\left(-\alpha \gamma\luv\ln10-\frac{\left[\lx-\gamma\luv-\beta\right]^2}{2\Sigx}\right)&
\nonumber\\&\propto\exp\left(-\alpha \lx\ln10-\frac{\left[\gamma\luv-\lx+\beta+\alpha\Sigx\ln10\right]^2}{2\Sigx}\right).&
\end{align}

We thus see that the probability density $f(\lx,\luv|\s{UV},\s{X},\gamma,\beta)$ for parameter values (1) $\gamma=\gamma_0$, $\beta=\beta_0$, $\s{UV}=0$, $\s{X}=\sigma_{\rm X0}$ is the same as for (2) $\gamma=\gamma_0$, $\beta=\beta_0+\alpha\sigma^2_{\rm X0}\ln10$, $\s{UV}=\sigma_{\rm X0}/\gamma_0$, $\s{X}=0$ [the right-hand sides of equations~(\ref{eq:fipluv}) and (\ref{eq:fiplx}) become identical]. 
This means that given a $f(\lx, \luv|\s{UV},\s{X},\gamma,\beta)$ distribution or a sample from it, it is impossible to give preference to one of these two sets of parameter values.
In other words, in the special case of a power-law LF and under assumption that the X-ray--UV luminosity relation is scattered in only one direction ($\lx$ or $\luv$), there arises degeneracy between the $\s{X}=0$ and $\s{UV}=0$ solutions. On the $\lx$ vs. $\luv$ plane, the primary dependence line of the $\s{X}=0$ solution is  translated by $\alpha\sigma^2_{\rm X0}\ln10$ along the X-ray axis from the  primary dependence line for the $\s{UV}=0$ solution.

However, this symmetry breaks down for any realistic LF, such as a double power-law LF used in this work. This allows us to give preference to either the $\s{X}=0$ or the $\s{UV}=0$ solution using e.g. the maximum likelihood method. Moreover, with the approach taken in this work we can consider the more general case where both $\s{X}$ and $\s{UV}$ are not equal to zero.

\section{Result for a higher optical flux threshold}
\label{appendix:E}

\begin{figure}[t]
    \begin{center}
    	\vspace{\wid}
    	\includegraphics[width=1\columnwidth]{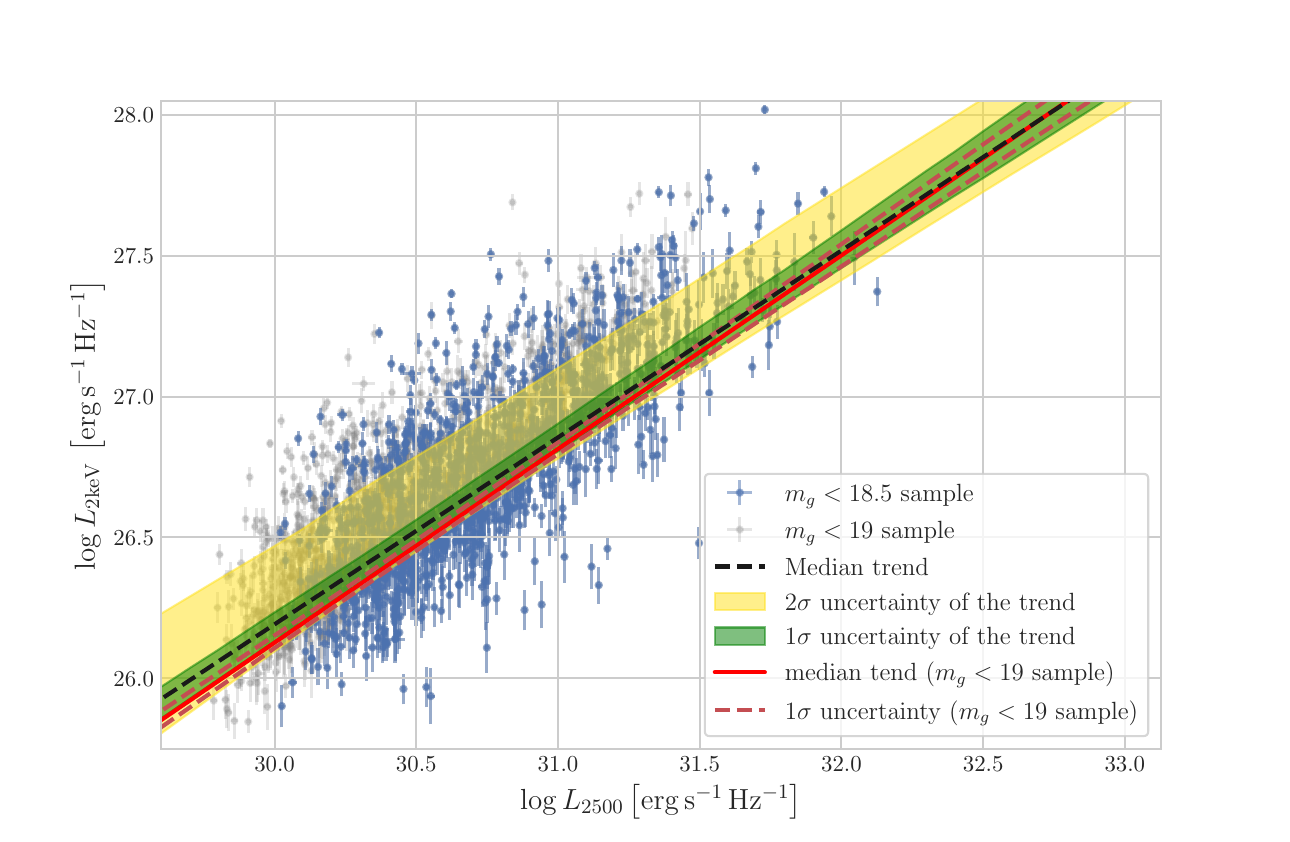}
    	\caption{Scatter plot of the X-ray and UV luminosities with their measurement uncertainties for the optically bright ($m_g<18.5$) quasar sample (points with error bars) and the inferred dependence between these quantities for the baseline model (the median trend and $1\sigma$ and $2\sigma$ uncertainty regions). For comparison, the red solid and red dashed lines show the median trend and $1\sigma$ uncertainty region for our main \srg/eROSITA--SDSS quasar sample ($m_g<19$).
        }
     \label{fig:resPLOT185} 
    \end{center}
\end{figure}

\begin{figure}[t]
    \begin{center}
    	\vspace{\wid}
    	\includegraphics[width=4in]{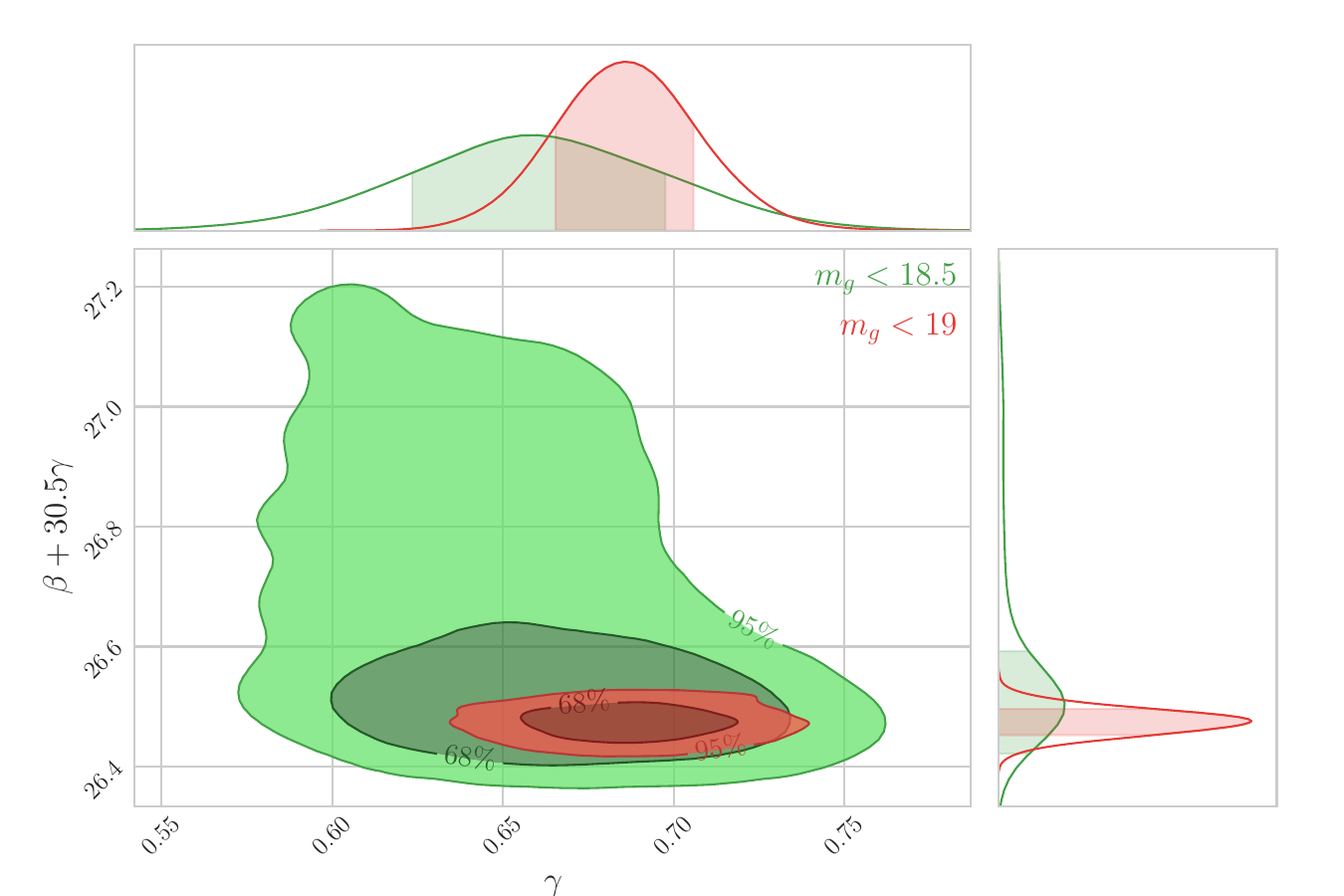}
    	\caption{Main panel: contour plots in the slope--normalization plane of the 68\% and 95\% smallest credible regions of the joint posterior distribution for the \srg/eROSITA--SDSS quasar sample ($m_g<19$, red color) and the optically bright quasar sample ($ m_g<18.5$, green color). Side panels: posterior distributions of the corresponding parameter for both samples and their 68\% smallest credible intervals (shaded regions).
        }
     \label{fig:contours} 
    \end{center}
\end{figure}

\begin{table*}[h]
	\vspace{2mm}
	\centering
	\vspace{2mm}
	\begin{tabular}{ccccccc} \hline
	Parameter & Mean value and & $2.5$\% percentile & 16\% percentile & median & 84\% percentile & $97.5$\% percentile\\ 
 & standard deviation & & & & & \\
 \hline
 \multicolumn{7}{c}{Baseline model}\\
 \hline

$\gamma$ & $0.66,\,0.04$ & $0.59$ & $0.62$ & $0.66$ & $0.70$ & $0.73$ \\
$\beta+30.5\gamma$ & $26.58,\,0.17$ & $26.45$ & $26.47$ & $26.51$ & $26.72$ & $27.07$ \\
$\sigma^2_{{\rm intUV}}$ & $0.03,\,0.04$ & $0.000$ & $0.001$ & $0.007$ & $0.062$ & $0.160$ \\
$\sigma^2_{{\rm intX}}$ & $0.054,\,0.018$ & $0.002$ & $0.040$ & $0.060$ & $0.068$ & $0.076$ \\
$g_1$ & $0.440,\,0.021$ & $0.40$ & $0.42$ & $0.44$ & $0.46$ & $0.48$ \\
$g_2$ & $4.2,\,0.3$ & $3.7$ & $3.9$ & $4.2$ & $4.5$ & $4.8$ \\
$l^*$ & $44.91,\,0.29$ & $44.25$ & $44.53$ & $45.05$ & $45.15$ & $45.21$ \\
$p_1$ & $3.34,\,0.08$ & $3.18$ & $3.26$ & $3.34$ & $3.43$ & $3.51$ \\
$z_c$ & $2.29,\,0.07$ & $2.15$ & $2.22$ & $2.29$ & $2.36$ & $2.44$ \\

 \hline
 \end{tabular}
   \caption{Characteristics of the posterior distributions of the model parameters based on the optically bright quasar sample.}
  \label{tab:parametersD} 
\end{table*}

To additionally test the robustness of the $\Lx$--$\Luv$ relation derived in this work, we repeated the entire analysis using a quasar sample with a higher optical flux threshold compared to our main \srg/eROSITA--SDSS quasar sample. Namely, we adopted $m_g<18.5$ instead of $m_g<19$, retaining the same X-ray flux limit. The resulting sample contains 1242 quasars. 

Figure~\ref{fig:resPLOT185} shows the $\Lx$ vs. $\Luv$ scatter-plot and the $\Lx (\Luv)$ primary dependence obtained by applying our baseline model (see Section~\ref{s:Baseline model}) to the optically bright sample, in comparison with the result for the main sample. Table~\ref{tab:parametersD} presents the posterior distribution characteristics of the model parameters for the optically bright sample. In addition, in Fig.~\ref{fig:contours} we compare the slope and normalization of the $\Lx (\Luv)$ dependencies inferred for the optically bright and main samples. Namely, we plot the joint posterior distribution contours corresponding to the 68\% and 95\% smallest credible regions. Note that the latter do not coincide with the quantile-based credible intervals listed in Table~\ref{tab:parametersD}, because the posterior distributions (especially for $\beta+30.5\gamma$) are highly skewed. 

The results obtained for the optically bright and main samples are in good agreement with each other. This strengthens the robustness of the conclusions of this work. It is important to note the $\Lx (\Luv)$ trend line passes approximately through the centre of the optically bright sample in Fig.~\ref{fig:resPLOT185}, which was not the case for our main sample. This explicitly demonstrates that our approach is insensitive to adopted X-ray and optical flux limits. 

\end{document}